\documentclass{aa}

\usepackage{natbib}
\usepackage{amsmath}
\allowdisplaybreaks
\usepackage{amssymb}
\usepackage{graphicx}
\usepackage{xcolor}
\usepackage[english]{babel}
\usepackage{hyperref}  
\hypersetup{colorlinks=true, citecolor=blue, linkcolor=blue}

\newcommand{\figpath}{Fig-eps/} 

\newcommand{\bea}{\begin{eqnarray}}
\newcommand{\eea}{\end{eqnarray}}

\begin{document}

\title{Modeling the evection resonance for trojan satellites: application to the Saturn system}
\titlerunning{Evection resonance in the trojan satellites of Saturn}
\author{Giuppone, C. A. \inst{1}, Roig, F.\inst{2} \and Saad-Olivera, X. \inst{2}}
\institute{Universidad Nacional de C\'ordoba, Observatorio Astron\'omico, IATE, Laprida 854, 5000 C\'ordoba, Argentina 
\and 
Observat\'orio Nacional, Rio de Janeiro, 20921-400, RJ, Brazil}

   \date{Received \dots; accepted \dots}
\abstract
{%
The stability of satellites in the solar system is affected by the so-called evection resonance. 
The moons of Saturn, in particular, exhibit a complex dynamical architecture in which co-orbital configurations occur, especially close to the planet where this resonance is present.
}{%
We address the dynamics of the evection resonance, with particular focus on the Saturn system, and compare the known behavior of the resonance for a single moon to that of a pair of moons in co-orbital trojan configuration.
}{%
We developed an analytic expansion of the averaged Hamiltonian of a trojan pair of bodies, including the perturbation from a distant massive body. {The analysis of the corresponding equilibrium points was restricted to the asymmetric apsidal corotation solution of the co-orbital dynamics.} We also performed numerical N-body simulations to construct dynamical maps of the stability of the evection resonance in the Saturn system, and to study the effects of this resonance under the migration of trojan moons caused by tidal dissipation.
}{%
The structure of the phase space of the evection resonance for trojan satellites is similar to that of a single satellite, differing in that the libration centers are displaced from their standard positions by an angle that depends on the periastron difference $\varpi_2-\varpi_1$ and on the mass ratio $m_2/m_1$ of the trojan pair. In the Saturn system, the inner evection resonance, located at $\sim 8\,R_\mathrm{S}$, may capture a pair of trojan moons by tidal migration; the stability of the captured system depends on the assumed values of the dissipation factor $Q$ of the moons. On the other hand, the outer evection resonance, located at $>0.4\,R_\mathrm{Hill}$, cannot exist at all for trojan moons, because trojan configurations are strongly unstable at distances from Saturn longer than $\sim 0.15\,R_\mathrm{Hill}$.
}{%
{The interaction with the inner evection resonance may have been relevant during the early evolution of the Saturn moons Tethys, Dione, and Rhea. In particular, Rhea may have had trojan companions in the past that were lost when it crossed the evection resonance, while Tethys and Dione may either have retained their trojans or have never crossed the evection. This may help to constrain the dynamical processes that led to the migration of these satellites and to the evection itself.}
}


\maketitle


\section{Introduction}
In studying the dynamics of satellites in the solar system, it is important to establish whether the bodies are bound to the planets or may escape to heliocentric orbits. Therefore, the perturbation by the Sun is a key issue when the stability of these satellites is to be addressed. In the case of the Moon, the development of Hill's lunar theory allowed identifying the so-called evection term \citep{brouwer_clemence_1961} as the largest periodic correction to the mean longitude of the Moon. For any satellite, the evection term is associated with the harmonic $\cos(2 \lambda_{\odot}-2 \varpi)$, where $\lambda_{\odot}$ is the mean longitude of the Sun and $\varpi$ is the longitude of the pericenter of the satellite. The evection resonance arises when the precession rate of $\varpi$ equals the solar mean motion, causing the angle $\lambda_{\odot}-\varpi$ to librate around an equilibrium point. In the classical evection, the pericenter precession is driven by the solar perturbation itself, but in principle, any other perturbation inducing a pericenter precession may originate an evection resonance.

In the restricted three-body problem, \citet{Henon_1969, Henon_1970} showed that the classical evection resonance appears as a bifurcation of a family of simple periodic orbits at a value of the semi-major axis $a = 0.45\, R_\mathrm{Hill}$.

\citet{Hamilton_1997} showed that in the case of prograde orbits, the evection resonance appears at $a = 0.53\, R_\mathrm{Hill}$ and is characterized by the resonant angle $\lambda_{\odot}- \varpi$ librating either around $0^{\circ}$ or $180^{\circ}$. This alignment or anti-alignment of the satellite pericenter with the Sun direction induces cumulative perturbations that may cause the escape of the satellite \citep{Nesvorny_etal_2003}, which might be relevant for the stability of bodies that are migrating as a result of tidal evolution or gas drag. 

The evection resonance plays an important role in sculpting the architecture of the satellite systems. \citet{Nesvorny_etal_2003} studied the orbital and collisional evolution of satellites and found that prograde satellite orbits with large semi-major axes are unstable because of the effect of the evection resonance. \citet{Cuk_2009} explored the fate of fictitious objects trapped in the lunar trojan points after the formation of the Moon, and found that these bodies can survive the tidal migration of the Moon until they reach 38 Earth radii, where the evection resonance ejects them from the system. \cite{Cuk_etal_2016} also studied the past evolution of the Tethys-Dione system and found that the evection resonance perturbing a pair of medium-size moons is the most likely mechanism for triggering instability and massive collisions in the Saturn system. Finally, \citet{Spalding_2016} studied the inward migration of giant exoplanets immersed in protoplanetary gas disks, and showed that dynamical interactions may naturally destroy the hypothetical moons of such planets through capture into the evection resonance. 
Several analytical approaches to the study of the evection resonance have been published. \citet{Yokoyama_etal_2008} analytically determined that the semi-major axis for the appearance of the evection resonance lies at $a = 0.529 R_\mathrm{Hill}$, while \cite{Frouard_2010} used dynamical maps to show that this resonance constitutes the outermost region of stability for the prograde satellites of Jupiter, being surrounded by chaotic orbits. The authors concluded that the evection can harbor stable orbits at distances from Jupiter of $\sim 0.43$ au.
\cite{Frouard_2010} also showed that the inclusion of third-order terms in the expansion of the solar disturbing function in Legendre polynomial causes the evection resonance to display a non-symmetric topology with respect to $\lambda_{\odot}- \varpi =0^{\circ}$ and $180^{\circ}$. 
\citet{Andrade-Ines_2018} constructed a secular theory for planetary motion around compact binaries and analyzed the occurrence of the evection resonance that is driven by the stellar companion. 
The study of the evection resonance using analytical models is not straightforward, however, because usually large perturbations act on the regions of interest. 
In this work, we study the occurrence and stability of the evection resonance in the satellite system of Saturn. This system is particularly interesting because it harbors the only known examples of co-orbital configurations in satellites in the solar system. These are 
\begin{enumerate}
\item the pair Janus and Epimetheus, which evolve in a relative horseshoe orbit \citep{murray_dermott_1999}, and may have originated from a former trojan configuration \citep{Gott_2005};
\item the trojan system of Tethys, accompanied by the moons Telesto (at L$_4$) and Calypso (at L$_5$) \citep[see][and references therein] {Niederman+2018};
\item the trojan system of Dione, accompanied by the moons Helene (at L$_4$) and Polydeuces (at L$_5$).
\end{enumerate}
Our main goal is to provide some clues about the possible past dynamics of both single and co-orbital moons. In Sect.~\ref{sec.analytical} we introduce an analytical expansion of the Hamiltonian for two satellites in trojan co-orbital configuration, and we apply it to describe the basic topology of the evection resonance in such a system. In Sect.~\ref{sec.methods} we introduce our numerical analysis of the stability of the evection resonance in the Saturn system, and we apply it first to the case of a single satellite (Sect.~\ref{sec.single}), and then to the case of a pair of trojan moons (Sect.~\ref{sec.trojans}). Section~\ref{tides} is devoted to studying the effect of the evection resonance on satellites that migrate as a result of tidal evolution. Finally, our conclusions are presented in Sect.~\ref{sec.conclusions}. 

\section{Analytical model for trojan satellites}\label{sec.analytical}
In this section, we set up a model to describe the evection
resonance in the case of two satellites in co-orbital trojan motion. 
We consider a hierarchical system of four bodies of masses $m_{i}$
($i=0,..,3$), in which $m_{0}$ is a central oblate body of radius
$R_{0}$ (planet), $m_{1},m_{2}\ll m_{0}$ are two co-orbital bodies
around $m_{0}$ (satellites), and $m_{3}$ is a distant perturber
(the Sun).  $\mathbf{u}_{i},\mathbf{w}_{i}$ are the barycentric
positions and momenta of the bodies, and we introduce the following
set of canonical coordinates:
\begin{align}
\mathbf{r}_{0} & =  \frac{1}{M}\sum_{j=0}^{3}m_{j}\mathbf{u}_{j}\nonumber \\
\mathbf{r}_{i} & =  \mathbf{u}_{i}-\mathbf{u}_{0}\qquad\qquad\qquad i=1,2\nonumber\\
\mathbf{r}_{3} & =  \mathbf{u}_{3}-\frac{1}{m}\sum_{j=0}^{2}m_{j}\mathbf{u}_{j},\label{eq:pos} 
\end{align}
where $m=\sum_{i=0}^{2}m_{i}$ and $M=m+m_{3}$. {The} positions $\mathbf{r}_{1},\mathbf{r}_{2}$ {are measured from} $m_{0}$
, while $\mathbf{r}_{3}$ {is measured from} the barycenter of the sub-system
$m_{0},m_{1},m_{2}$. Their conjugated momenta are
\begin{align}
\mathbf{p}_{0} & =  \sum_{j=0}^{3}\mathbf{w}_{j}\nonumber \\
\mathbf{p}_{i} & =  \mathbf{w}_{i}-\frac{m_{i}}{m}\sum_{j=0}^{2}\mathbf{w}_{j}\qquad\qquad i=1,2\nonumber\\
\mathbf{p}_{3} & =  \mathbf{w}_{3}-\frac{m_{3}}{M}\sum_{j=0}^{3}\mathbf{w}_{j}.\label{eq:mom} 
\end{align}
Defining
\begin{eqnarray}
\mu_{i}=\frac{m_{i}m_{0}}{m_{i}+m_{0}} & \qquad & \beta_{i}=G(m_{0}+m_{i})\qquad i=1,2\nonumber \\
\mu_{3}=\frac{m_{3}m}{m_{3}+m} & \qquad & \beta_{3}=GM,\end{eqnarray}
the Hamiltonian of the system reads
\begin{equation}
\mathcal{H}=\sum_{i=1}^{3}\left(\frac{p_{i}^{2}}{2\mu_{i}}-\frac{\beta_{i}\mu_{i}}{r_{i}}\right)+\sum_{i=1}^{2}\mathcal{J}_{i}+\mathcal{R}_{12}+\mathcal{R}_{3},\label{eq:hamt}
\end{equation}
where $\mathcal{J}_{i}$ is the oblateness
perturbation felt by $m_{i}$,
\begin{equation}
\mathcal{R}_{12}  = -\frac{Gm_{1}m_{2}}{\left|\mathbf{r}_{1}-\mathbf{r}_{2}\right|}+\frac{1}{m_{0}}\mathbf{p}_{1}\cdot\mathbf{p}_{2}
\end{equation}
is the co-orbital disturbing function, and
\begin{align}
\mathcal{R}_{3} & = Gm_{3}m_{0}\left(\frac{1}{r_{3}}-\frac{1}{r_{3}^{\prime}}\right)\nonumber\\
& +Gm_{3}\sum_{i=1}^{2}m_{i}\left(\frac{1}{r_{3}}-\frac{1}{\left|\mathbf{r}_{3}^{\prime}-\mathbf{r}_{i}\right|}\right)\label{eq:distpert}
\end{align}
is the distant perturbation, with
the auxiliary variable
\begin{equation}
\mathbf{r}_{3}^{\prime}=\mathbf{r}_{3}+\frac{1}{m}\sum_{i=1}^{2}m_{i}\mathbf{r}_{i}\label{eq:rbary}
,\end{equation}
which is the position of $m_{3}$ relative to $m_{0}$ \citep{Chambers2010}. 

Assuming that $R_{0}\ll r_{i}$ $(i=1,2)$, we may expand $\mathcal{J}_{i}$ in powers of $R_{0}/r_{i}$ using Legendre polynomials. Up to the fourth order, we obtain
\begin{align}
\mathcal{J}_{i} & = \beta_{i}\mu_{i}J_{2}\frac{R_{0}^{2}}{r_{i}^{3}}\left(\frac{3}{2}\sin^{2}\varphi_{i}-\frac{1}{2}\right)\nonumber\\
& +\beta_{i}\mu_{i}J_{4}\frac{R_{0}^{4}}{r_{i}^{5}}
\left(\frac{35}{8}\sin^{4}\varphi_{i}-\frac{30}{8}\sin^{2}\varphi_{i}+\frac{3}{8}\right)+\ldots\label{eq:hamt2d}
\end{align}
where $J_{k}$ {are} the zonal harmonic {coefficients} of $m_{0}$, and $\varphi_{i}$
is the latitude of $m_{i}$ over the $m_{0}$ equator \citep{murray_dermott_1999}.

Similarly, assuming that $\mathbf{r}_{i}\ll\mathbf{r}_{3}$
$(i=1,2)$, we may expand $\mathcal{R}_{3}$ in powers of $r_{i}/r_{3}$. Taking into account Eq. (\ref{eq:rbary}) and introducing the small quantities
\begin{equation}
\bar{\nu}=\frac{m_{1}m_{2}}{m}\qquad\qquad\nu_{i}=\frac{m_{i}}{m},\qquad i=1,2\label{eq:nu}
,\end{equation}
we obtain up to the third order\footnote{%
In most satellite systems, the ratio $R_0/r_i$ is usually a few orders of magnitude higher than the ratio $r_i/r_3$, therefore it makes sense to carry on the expansion of $\mathcal{J}_i$ up to the fourth order while truncating the expansion of $\mathcal{R}_3$ at the third order.}
\begin{equation}
\mathcal{R}_{3} = \mathcal{R}_{3}^{(2)}+\mathcal{R}_{3}^{(2c)}+\mathcal{R}_{3}^{(3)}+\mathcal{R}_{3}^{(3c)}+\ldots\label{eq:hamt2c}
,\end{equation}
with
\begin{align}
\mathcal{R}_{3}^{(2)} & = -Gm_{3}\sum_{i=1}^{2}m_{i}\left(1-\nu_{i}\right)\frac{r_{i}^{2}}{r_{3}^{3}}\left(\frac{3}{2}\cos^{2}\gamma_{i}-\frac{1}{2}\right)\\
\mathcal{R}_{3}^{(2c)} & = Gm_{3}\,\bar{\nu}\frac{r_{1}r_{2}}{r_{3}^{3}}\left(3\cos\gamma_{1}\cos\gamma_{2}-\cos\phi\right)\\
\mathcal{R}_{3}^{(3)} & = -Gm_{3}\sum_{i=1}^{2}m_{i}\left(1-3\nu_{i}+2\nu_{i}^{2}\right)\nonumber\\
& \qquad\qquad\qquad\qquad\times\frac{r_{i}^{3}}{r_{3}^{4}}\left(\frac{5}{2}\cos^{3}\gamma_{i}-\frac{3}{2}\cos\gamma_{i}\right)\\
\mathcal{R}_{3}^{(3c)} & = Gm_{3}\,\bar{\nu}\left(1-2\nu_{1}\right)\frac{r_{1}^{2}r_{2}}{r_{3}^{4}}\left(\frac{15}{2}\cos^{2}\gamma_{1}\cos\gamma_{2}\right.\nonumber\\
& \qquad\qquad\qquad\qquad\quad\left.-\frac{3}{2}\cos\gamma_{2}-3\cos\gamma_{1}\cos\phi\right)\nonumber\\
& +Gm_{3}\,\bar{\nu}\left(1-2\nu_{2}\right)\frac{r_{1}r_{2}^{2}}{r_{3}^{4}}\left(\frac{15}{2}\cos^{2}\gamma_{2}\cos\gamma_{1}\right.\nonumber\\
& \qquad\qquad\qquad\qquad\quad\left.-\frac{3}{2}\cos\gamma_{1}-3\cos\gamma_{2}\cos\phi\right)
\end{align}
where $\gamma_{i}$ is the angle between $\mathbf{r}_{3}$ and $\mathbf{r}_{i}$, and
$\phi$ is the angle between $\mathbf{r}_{1}$ and $\mathbf{r}_{2}$. The terms $\mathcal{R}_{3}^{(ic)}$ arise from the motion of the co-orbital pair. When either $m_1$ or $m_2$ is zero, then $\mathcal{R}_{3}^{(ic)}=0$ and Eq. (\ref{eq:hamt2c}) reduces to the classical expansion of the lunar theory \citep{brouwer_clemence_1961}. 

No expansion is made of the co-orbital perturbation, since $r_{1}\sim r_{2}$, but taking into account that the osculating velocity of $m_{i}$ is
\begin{align}
\mathbf{v}_{i}=\frac{\mathbf{p}_{i}}{\mu_{i}}=\frac{m_{0}+m_{i}}{m_{0}m}\left[(m_{0}+m_{j})\dot{\mathbf{r}}_{i}-m_{j}\dot{\mathbf{r}}_{j}\right]\qquad & i=1,2 \nonumber\\
& j\neq i,\label{eq:velocity-1}
\end{align}
we may write
\begin{align}
\mathcal{R}_{12} & = -Gm_{1}m_{2}\left(r_{1}^{2}+r_{2}^{2}-2r_{1}r_{2}\cos\phi\right)^{-1/2}\nonumber \\
 & +\bar{\nu}\left(1+\frac{2\bar{\nu}}{m_{0}}\right)\dot{r}_{1}\dot{r}_{2}\cos\phi^{\prime}-\frac{\bar{\nu}^{2}}{\mu_{1}\mu_{2}}\left(\mu_{1}\dot{r}_{1}^{2}+\mu_{2}\dot{r}_{2}^{2}\right)\label{eq:hamt2b}
\end{align}
where $\phi^{\prime}$ is the angle between $\dot{\mathbf{r}}_{1}$ and $\dot{\mathbf{r}}_{2}$.

Equations (\ref{eq:hamt2d}), (\ref{eq:hamt2c}), and (\ref{eq:hamt2b})
can be further expanded in terms of the orbital elements of the Keplerian
motion. We restrict the motion to
coplanar orbits and let the equator of $m_{0}$ coincide with the
reference plane ($\varphi_i=0$). We introduce the set of canonical elements
\begin{eqnarray}
\theta_{1}=\frac{\lambda_{2}-\lambda_{1}}{2}; &  & Z_{1}=\mu_{2}\sqrt{\beta_{2}a_{2}}-\mu_{1}\sqrt{\beta_{1}a_{1}}\nonumber \\
\theta_{2}=\frac{\lambda_{2}+\lambda_{1}}{2}; &  & Z_{2}=\mu_{2}\sqrt{\beta_{2}a_{2}}+\mu_{1}\sqrt{\beta_{1}a_{1}}\nonumber \\
\theta_{3}=\lambda_{3}; &  & Z_{3}=\mu_{3}\sqrt{\beta_{3}a_{3}}-W_{1}-W_{2}\nonumber \\
\psi_{i}=\lambda_{3}-\varpi_{i}; &  & W_{i}=\mu_{i}\sqrt{\beta_{i}a_{i}}\left(1-\sqrt{1-e_{i}^{2}}\right),\;\; i=1,2\nonumber \\
\psi_{3}=-\varpi_{3}; & & W_{3}=\mu_{3}\sqrt{\beta_{3}a_{3}}\left(1-\sqrt{1-e_{3}^{2}}\right)\label{eq:resovar}
,\end{eqnarray}
where $a_{i},e_{i},\lambda_{i},\text{and }\varpi_{i}$ are the semi-major axis,
eccentricity, mean longitude, and pericenter longitude, respectively,
and we expand the Hamiltonian in powers of the eccentricities, around
$e_{i}=0$, up to $\mathcal{O}(e^4)$. 

The co-orbital motion is
characterized by the libration of the resonant angle $\theta_{1}$, while the synodic
angle $\theta_{2}$ is a fast angle that can be eliminated by an averaging:
\begin{equation}
\bar{\mathcal{H}}=\frac{1}{2\pi}\int_{0}^{2\pi}\mathcal{H}\,d\theta_{2}\label{eq:aver}
,\end{equation}
thus $Z_{2}$ becomes an integral of motion. 

In this way,
the averaged co-orbital perturbation, $\bar{\mathcal{R}}_{12}$, reduces
to an expansion similar to that obtained by \citet{Robutel_2013}. This
expansion has a singularity when $a_{1}=a_{2}$ and $\theta_{1}=0$,
even if $e_{i}\neq0$, and only contains terms of degree $1,e^2,e^4$, since the odd-degree terms are eliminated by the average. The averaged expansion of the oblateness perturbation, $\bar{\mathcal{J}}_{i}$, is
straightforward and also contains terms of degree $1,e^2,e^4$. Only
the averaged expansion of the distant perturbation, $\bar{\mathcal{R}}_{3}$, contains both even- and odd-degree terms in $e$. In particular, $\bar{\mathcal{R}}_{3}^{(2)}$ only contains terms of degree $1,e^2$, while $\bar{\mathcal{R}}_{3}^{(2c)}$ contains terms of degree $1,e^2,e^4$. Both $\bar{\mathcal{R}}_{3}^{(3)}$ and $\bar{\mathcal{R}}_{3}^{(3c)}$ contain terms of degree $e,e^3$ only.

After rearranging terms, the averaged Hamiltonian may be written as
\begin{align}
\bar{\mathcal{H}} & = \bar{\mathcal{H}}_{0}\left(\theta_{1},Z_{1},Z_{2},Z{}_{3}\right)\nonumber\\
& +\bar{\mathcal{H}}_{1}\left(\theta_{1},\psi_{1},\psi_{2},Z_{1},Z_{2},Z_{3},W_{1},W_{2}\right)\nonumber \\
& +\bar{\mathcal{H}}_{2}\left(\theta_{1},\theta_{3},\psi_{1},\psi_{2},\psi_{3},Z_{1},Z_{2},Z_{3},W_{1},W_{2},W_{3}\right)\label{eq:haver}
\end{align}
where $\bar{\mathcal{H}}_{0}$ is a one degree of freedom Hamiltonian
that contains all the terms that do not depend on the eccentricities,
$\bar{\mathcal{H}}_{1}$ is a perturbation that contains the terms that depend solely on $e_1,e_2$, and $\bar{\mathcal{H}}_{2}$ is a perturbation that contains all the
terms depending on $e_{3}$. Elimination of two degrees of freedom
is immediate by setting $e_{3}=0$. In that case, $\bar{\mathcal{H}}_{2}$
vanishes and the resulting Hamiltonian does not depend on $W_{3},\psi_{3}$,
nor on $\theta_{3}$, thus $Z_{3}$ becomes an integral of motion. This
integral reflects the perturbation of the co-orbital system onto the
distant perturber, which in this approach has a fixed circular orbit
with a varying semi-major axis. 

\begin{figure*}
\centering{}\includegraphics[height=5.5cm]{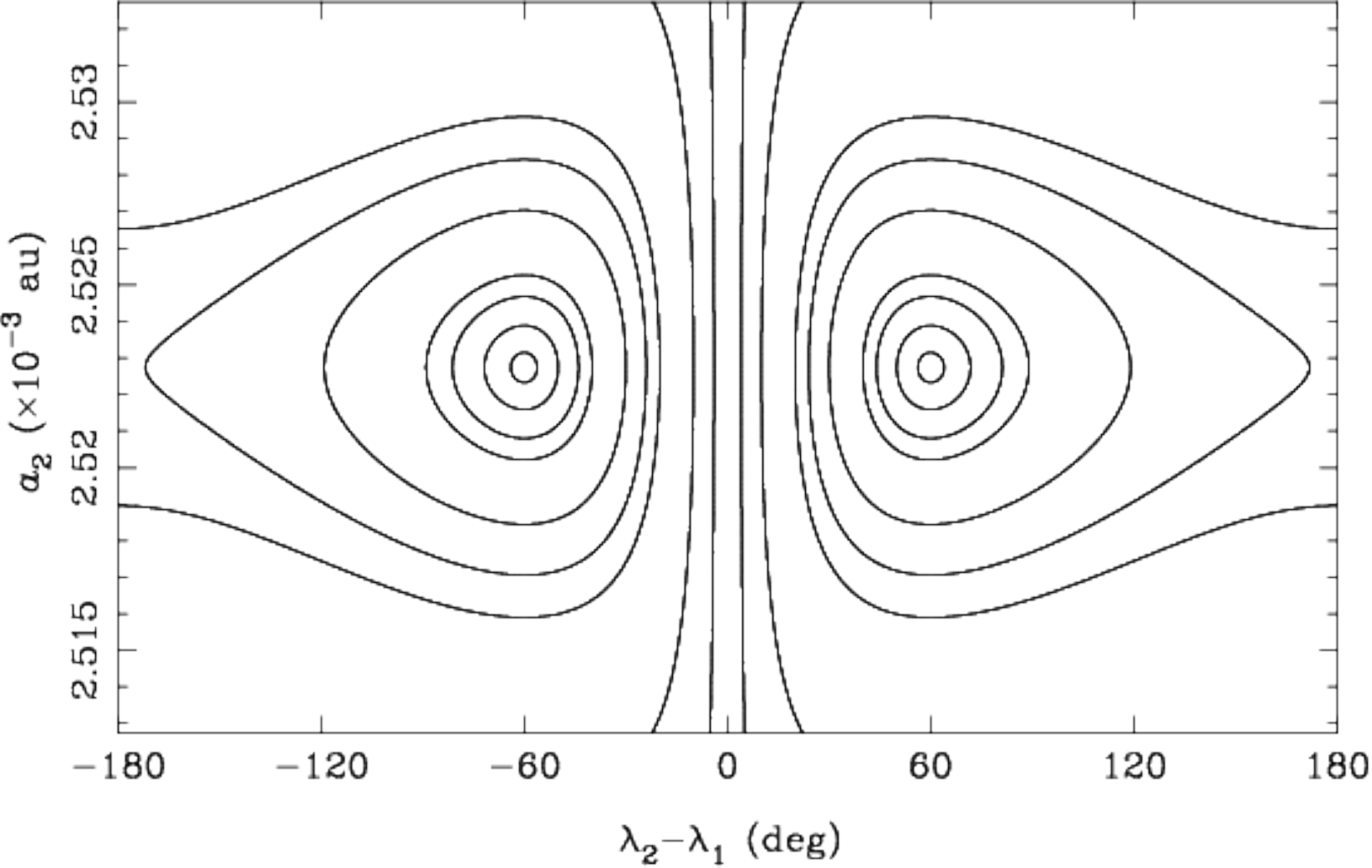}\;\;\;\includegraphics[height=5.5cm]{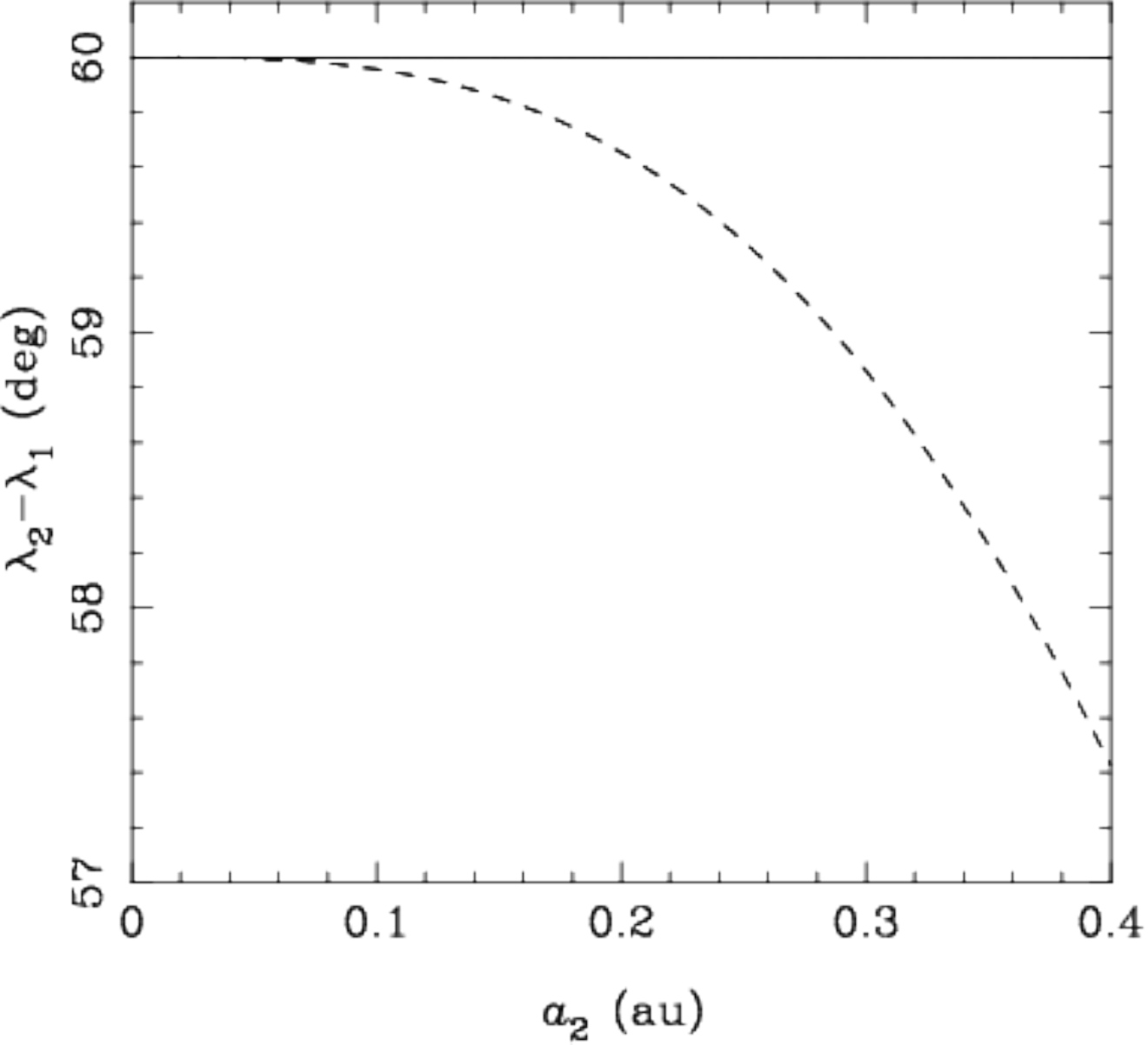}
\caption{\textit{Left.} Phase space of the Hamiltonian $\bar{\mathcal{H}}_{0}$
for the Dione-Helene system. \textit{Right.} Shift
of the trailing libration center (dashed line) in our model with respect
to the simple co-orbital motion (full line).}
\label{dione-helene-h0}
\end{figure*}

\begin{figure}
\centering                                     
\includegraphics[width=0.6\columnwidth]{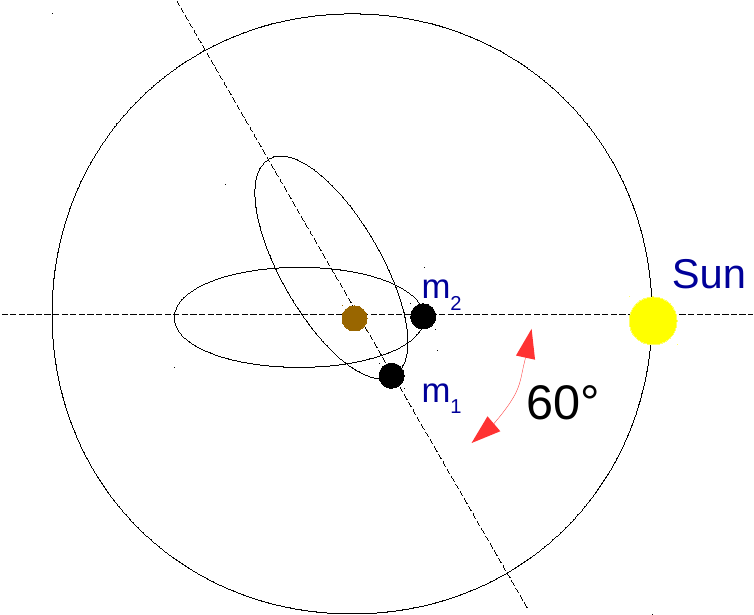}
\caption{Schematic orientation of the orbits for the trojan satellites in the ACR condition, assuming $\lambda_{3} = 0^{\circ}$ and $\psi_2=0^\circ$.}
\label{fig-orientation}

\end{figure}
  
The literal expansion of $\bar{\mathcal{H}}$ in non-canonical elements 
contains more than 50 terms, but we can restrict it to those of the lowest degree and to the most relevant terms. For the distant perturbation, we restrict the terms arising from the co-orbital motion, $\bar{\mathcal{R}}_{3}^{(2c)}$ and $\bar{\mathcal{R}}_{3}^{(3c)}$, to degree $e^2$, and keep the main terms, $\bar{\mathcal{R}}_{3}^{(2)}$ and $\bar{\mathcal{R}}_{3}^{(3)}$, up to degree $e^3$. This is a reasonable assumption when we take into account that $\nu_i \protect\la \mathcal{O}(r_i/r_3)$ for most satellite systems. Therefore, the term $\bar{\mathcal{R}}_{3}^{(3c)}$ is $\mathcal{O}(\nu_i)$ smaller than the term $\bar{\mathcal{R}}_{3}^{(3)}$, while the term $\bar{\mathcal{R}}_{3}^{(2c)}$ is on the same order as $\bar{\mathcal{R}}_{3}^{(3)}$. For the co-orbital perturbation, $\bar{\mathcal{R}}_{12}$, we follow \citet{Robutel_2013} and keep only terms of degree $e^2$. Finally, for the oblateness perturbation, we assume $J_4=0$. The averaged Hamiltonian then reduces to the following relevant terms:
\begin{align}
\bar{\mathcal{H}} & =-\sum_{i=1}^{3}\frac{\beta_{i}\mu_{i}}{2a_{i}}+\sum_{i=1}^{2}\left(\frac{A_{0i}}{a_{3}^{3}}+D_{i}\right)+B_{0}+\frac{C_{0}}{a_{3}^{3}}\nonumber\\
 & +\sum_{i=1}^{2}3\frac{A_{1i}}{a_{3}^{4}}e_{i}\cos\psi_{i}+\sum_{i=1}^{2}\frac{3}{2}\left(\frac{A_{0i}}{a_{3}^{3}}+D_{i}\right)e_{i}^{2}\nonumber\\
 & +\sum_{i=1}^{2}\left(B_{1}-\frac{1}{2}\frac{C_{0}}{a_{3}^{3}}\right)e_{i}^{2}+\sum_{i=1}^{2}\frac{15}{8}D_{i}e_{i}^{4}\nonumber\\
 & +\sum_{i=1}^{2}\frac{3}{2}\left(5\frac{A_{0i}}{a_{3}^{3}}+\frac{1}{4}\frac{C_{0}}{a_{3}^{3}}\right)e_{i}^{2}\cos2\psi_{i}\nonumber\\
 & +\left(B_{2}+\frac{C_{1}+C_{2}}{a_{3}^{3}}\right)e_{1}e_{2}\cos(\psi_{1}-\psi_{2})\nonumber\\
 & +3\frac{C_{2}}{a_{3}^{3}}e_{1}e_{2}\cos(\psi_{1}+\psi_{2})\nonumber \\
 & +\left(B_{3}+\frac{C_{3}}{a_{3}^{3}}\right)e_{1}e_{2}\sin(\psi_{1}-\psi_{2})\nonumber\\
 & +\frac{C_{4}}{a_{3}^{3}}\left(e_{1}^{2}\sin2\psi_{1}-e_{2}^{2}\sin2\psi_{2}\right)\nonumber\\
  & +\sum_{i=1}^{2}\frac{1}{4}\frac{A_{1i}}{a_{3}^{4}}e_{i}^{3}\left(9\cos\psi_{i}+35\cos3\psi_{i}\right)
\label{eq:hevec}
\end{align}
where $A_{ki},B_{k},C_{k}$ , and $D_{i}$ are functions of $\theta_{1},Z_{1},\text{and }Z_{2}$. Expressions for these functions in terms of non-canonical elements are given in the Appendix.

We note that when $m_{3}=0$, all the $A_{ki}$ and $C_{k}$ functions
are zero and the expansion reduces to the same expansion as reported by \citet{Robutel_2013}, with the addition of the oblateness terms $D_{i}$.
On the other hand, when $m_{2}=0$, all the $B_{k}$ and $C_{k}$
functions are zero, and the expansion reduces to the classical expansion
of the lunar theory, again with the addition of the oblateness terms \citep{Frouard_2010}.

For $e_1=e_2=0$, the Hamiltonian 
\begin{align}
\bar{\mathcal{H}}_{0} & =-\sum_{i=1}^{3}\frac{\beta_{i}\mu_{i}}{2a_{i}}+\sum_{i=1}^{2}\left(\frac{A_{0i}}{a_{3}^{3}}+D_{i}\right)+B_{0}+\frac{C_{0}}{a_{3}^{3}}
\label{eq:H0}
\end{align}
reflects the zeroth-order
topology of the co-orbital motion. This is shown in Fig.~\ref{dione-helene-h0}~(\textit{left})
for the Dione-Helene system. We note that at variance with the model
of \citet{Robutel_2013}, the inclusion in $\bar{\mathcal{H}}_{0}$
of the zeroth-order terms of the solar perturbation ($A_{0i},C_{0}$)
and the planet oblateness ($D_{i}$) shift the libration center
to values lower than $60^{\circ}$ at increasing distances from
the planet, as shown in Fig.~\ref{dione-helene-h0}~(\textit{right}).

\begin{figure*}
\centering{}\includegraphics[width=0.23\textwidth]{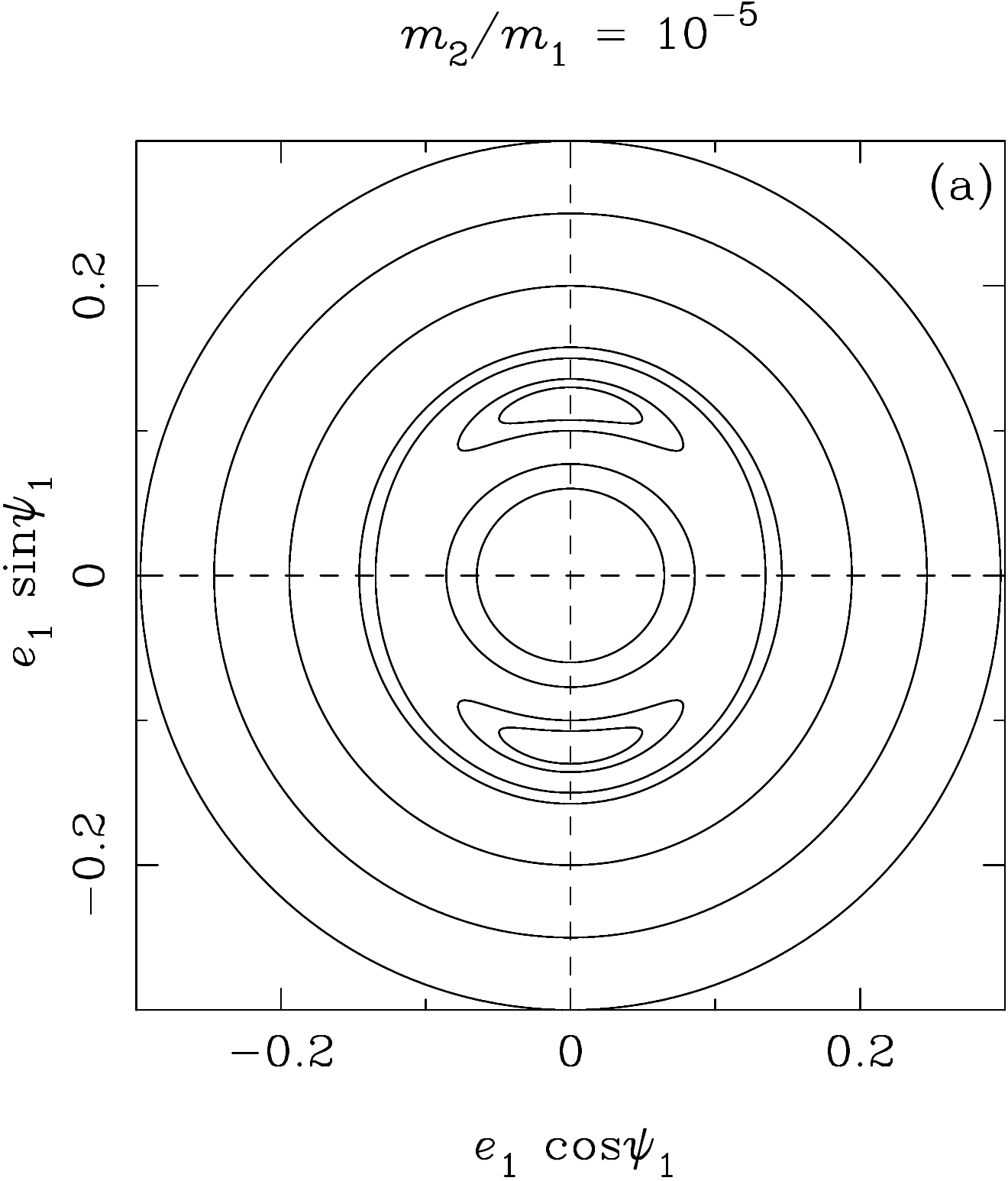}\;\;\;\includegraphics[width=0.23\textwidth]{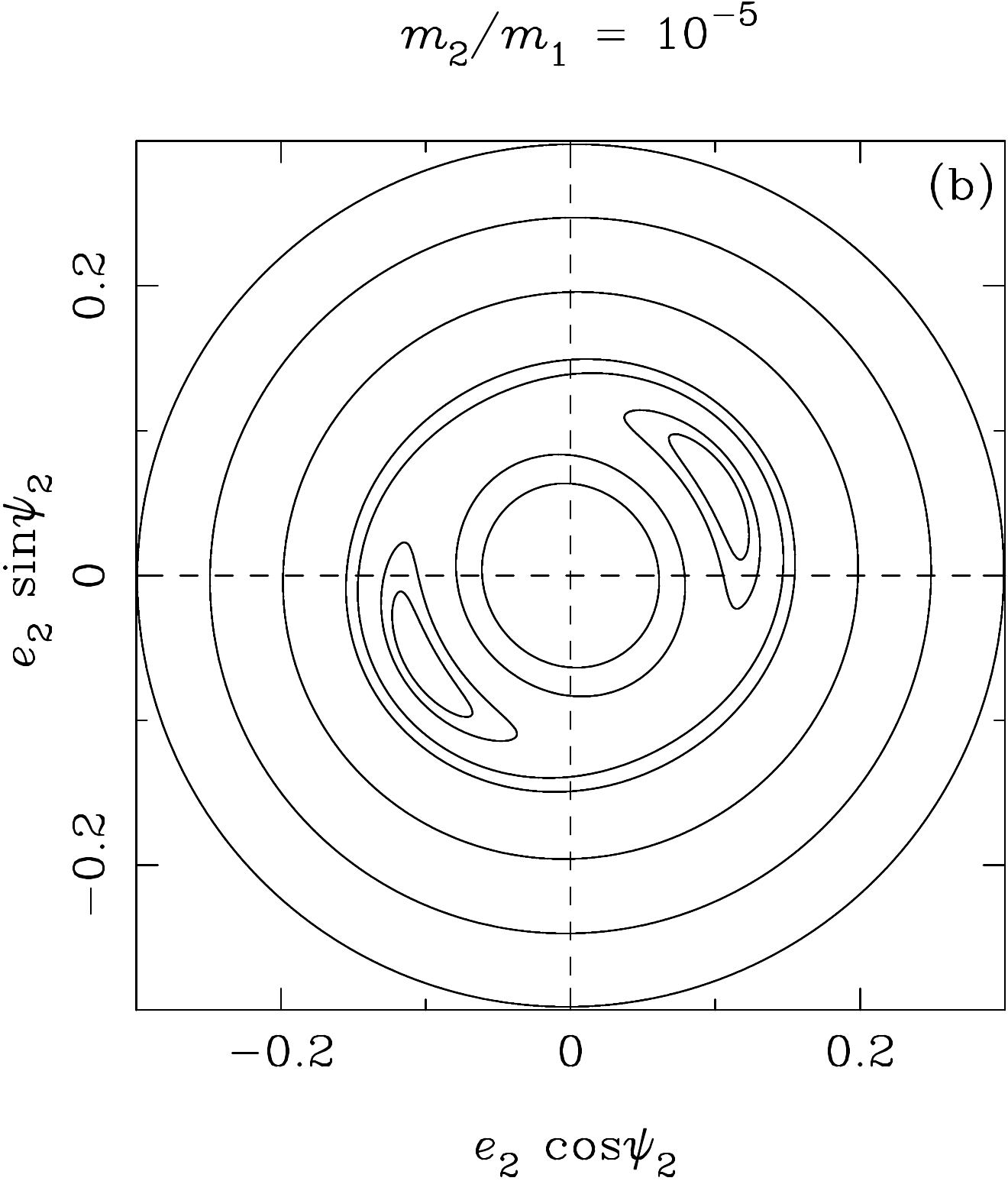}\;\;\;\includegraphics[width=0.23\textwidth]{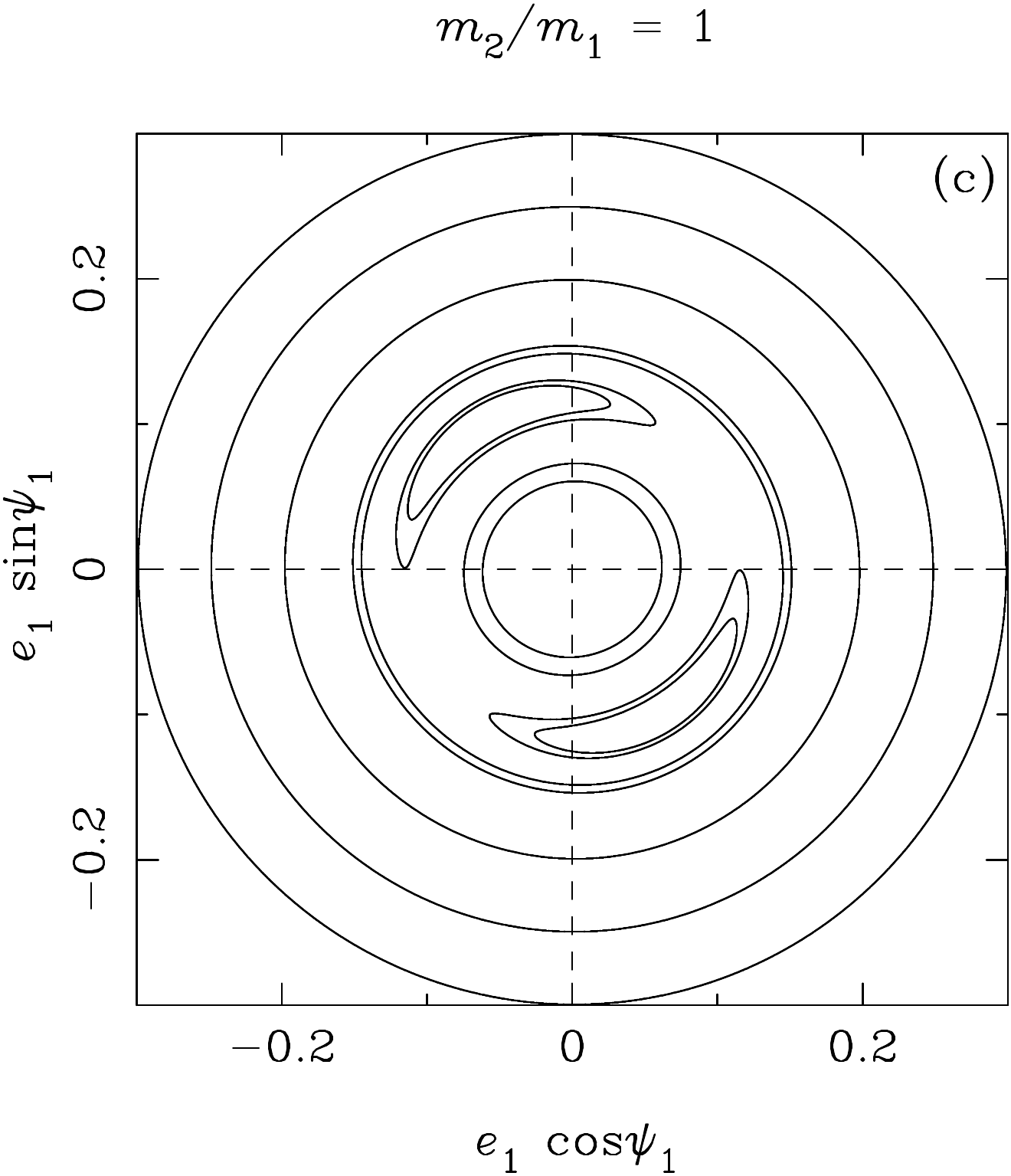}\;\;\;\includegraphics[width=0.23\textwidth]{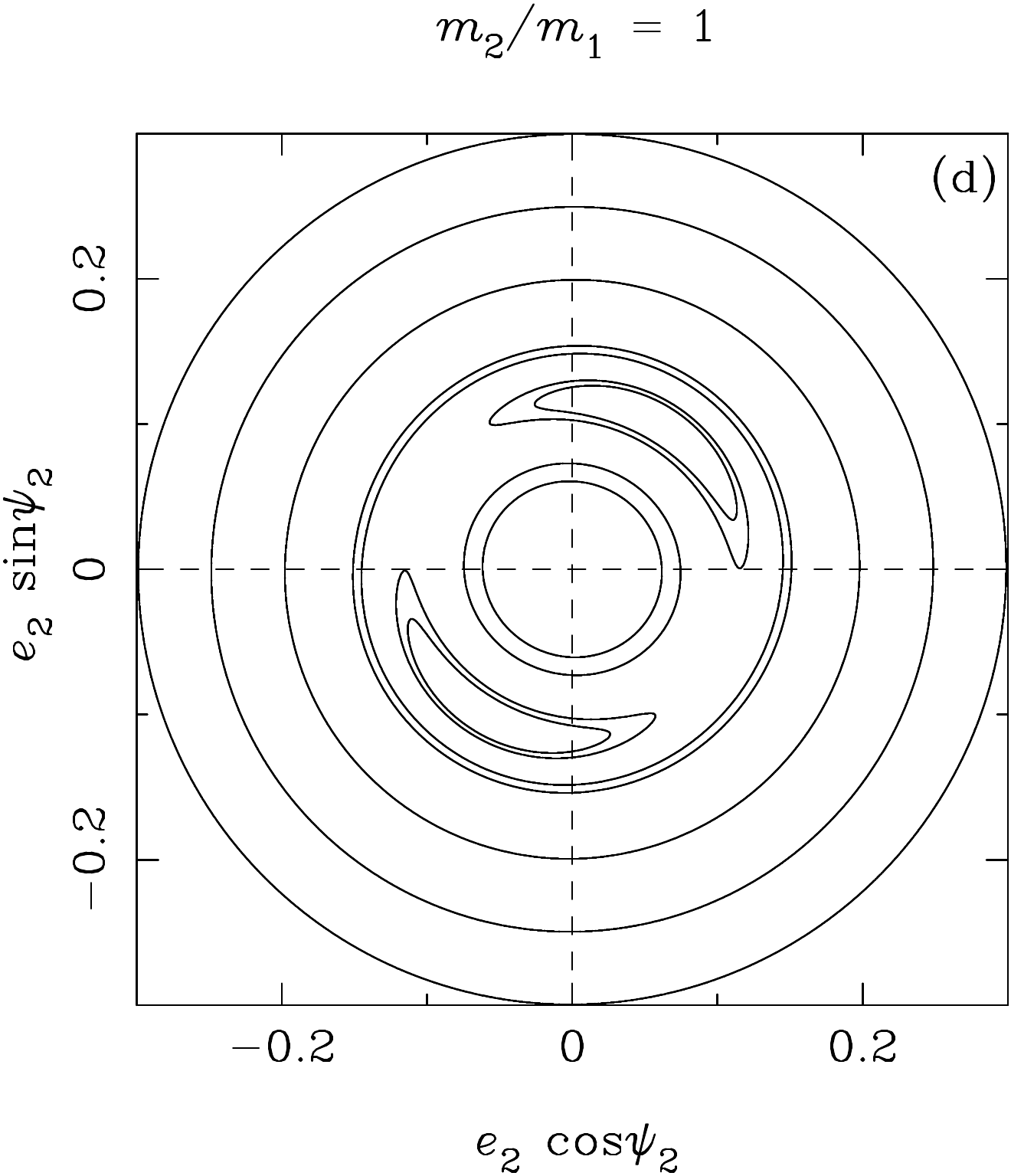}
\caption{Topology of the inner evection resonance for different mass ratios. Panels (a) and (b)  approximately correspond to the Dione-Helene pair. In all cases, $a_{1}=a_{2}=0.00326$ au (488~000 km) and $\Delta\varpi =\varpi_{2}-\varpi_{1}=60^{\circ}$.}
\label{topoin}
\end{figure*}

\begin{figure*}
\centering{}\includegraphics[width=0.31\textwidth]{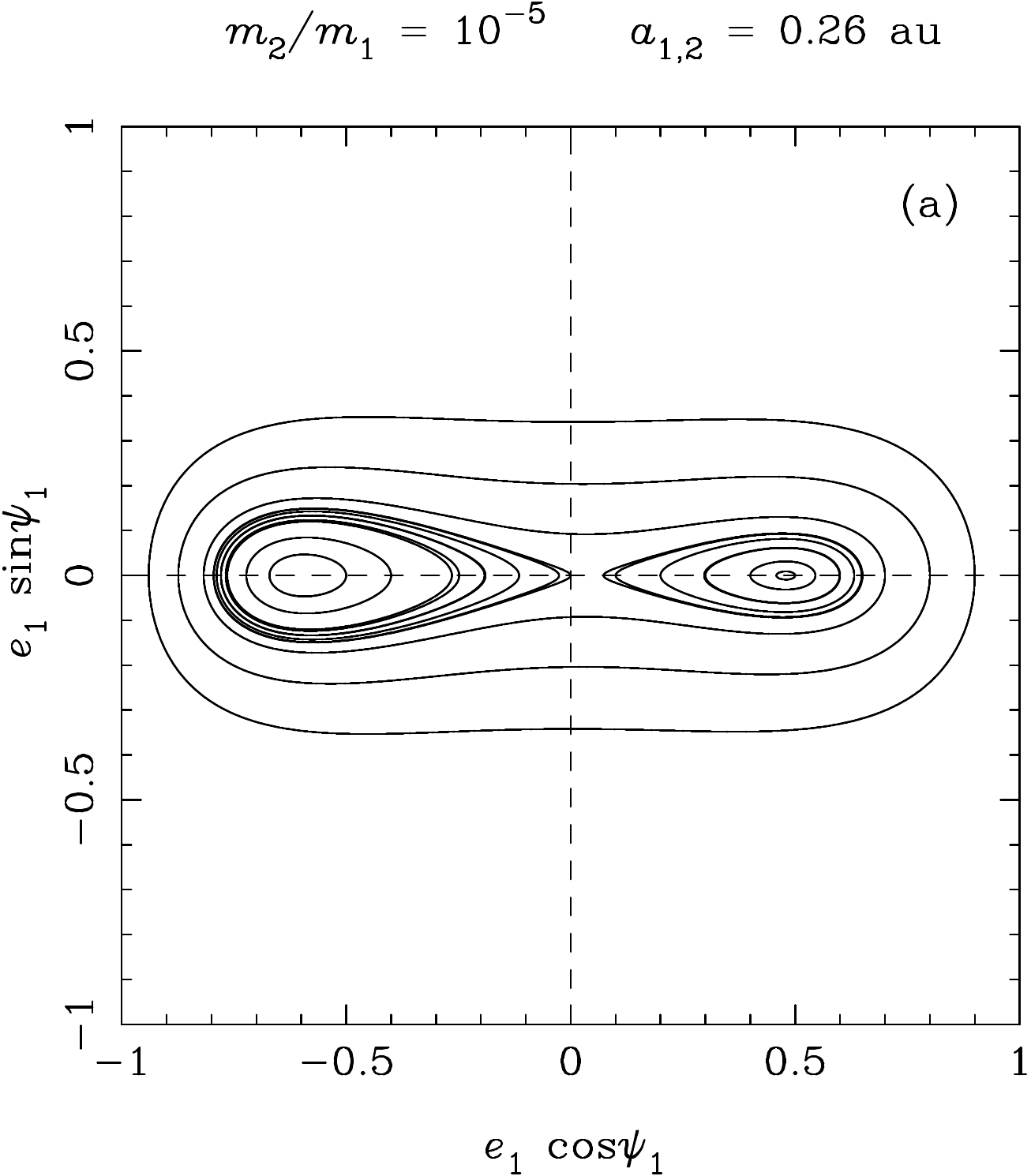}
\includegraphics[width=0.31\textwidth]{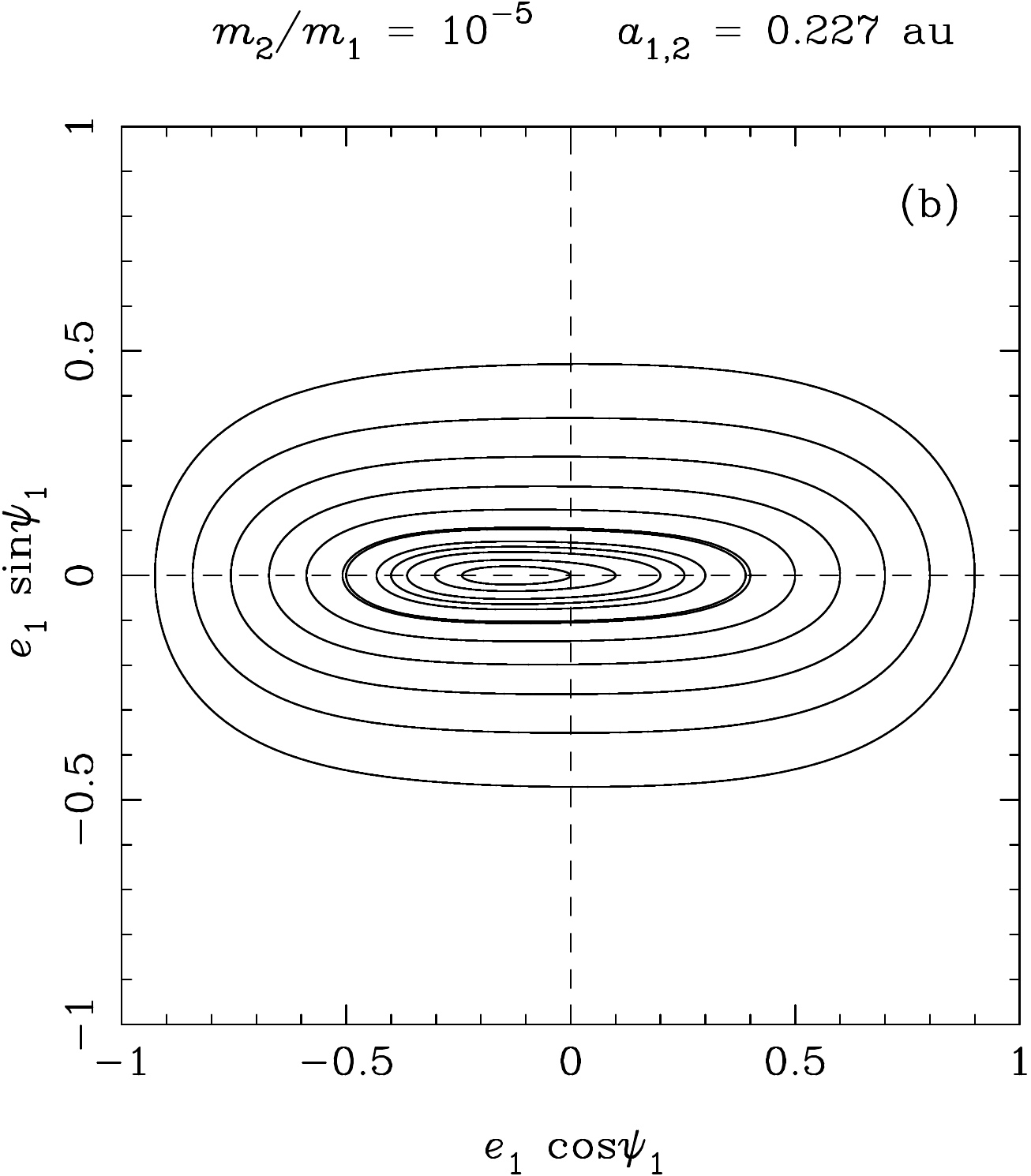}
\includegraphics[width=0.31\textwidth]{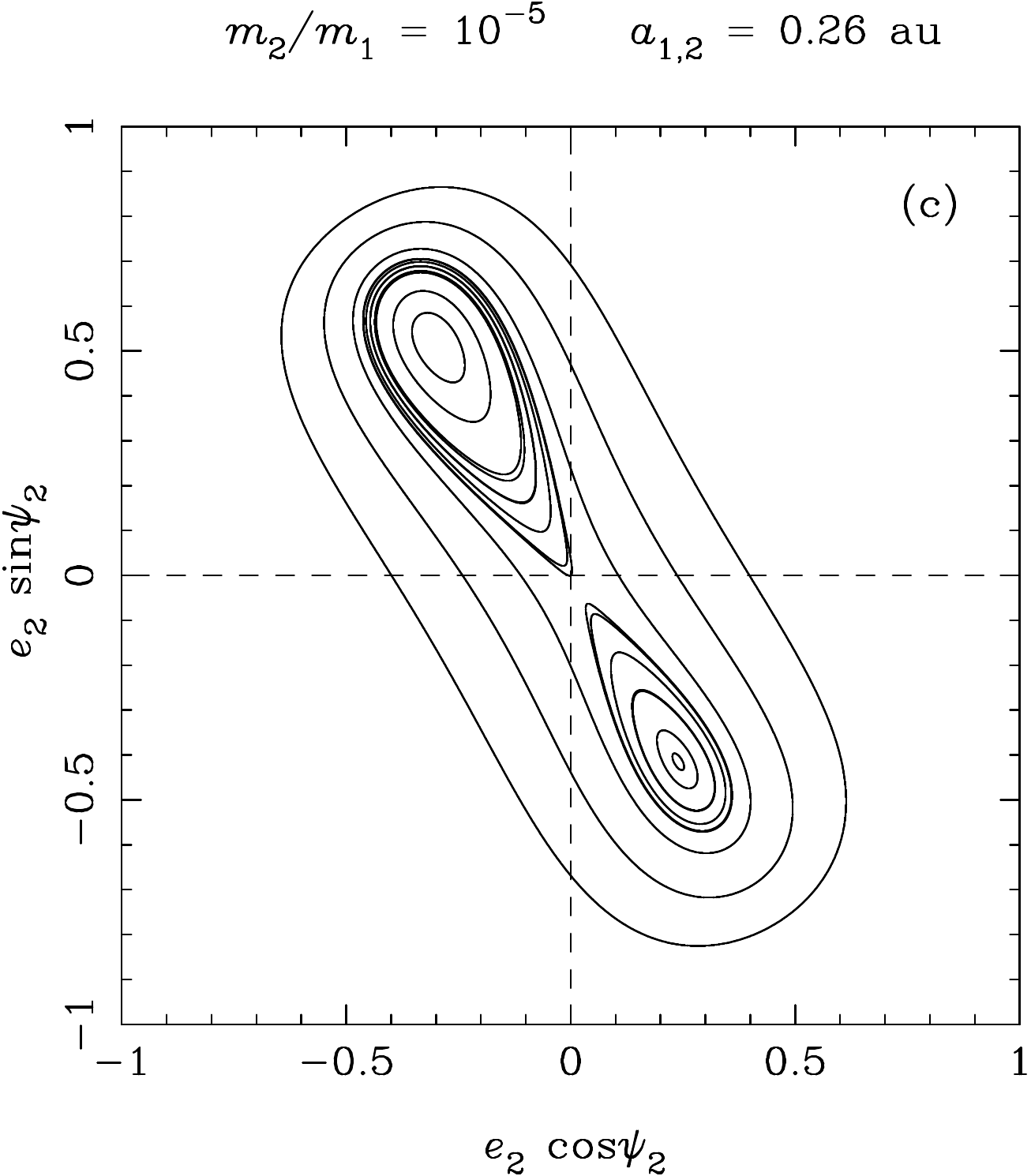}
\caption{Topology of the outer evection resonance for $m_{2}\ll m_{1}$.
Panels (a) and (b) show the behavior of $m_{1}$ at two different distances from the planet (in all cases $a_1=a_2$). In panel (a), two libration islands at $\psi_{1}=0^{\circ}$ and $180^{\circ}$ are connected by an unstable point that is shifted to positive values of $e_{1}\cos \psi_{1}$. In panel (b) only a kinematic libration around $\psi_{1}=180^{\circ}$ is observed. Panel (c) shows the behavior of $m_{2}$ at the same distance from the planet as in panel (a). In this case, the topology is the same, but the libration islands are rotated by $-\Delta\varpi$. In all cases, $\Delta\varpi = 60^{\circ}$.}
\label{topout}
\end{figure*}

\begin{figure}
\centering{}\includegraphics[width=1.0\columnwidth]{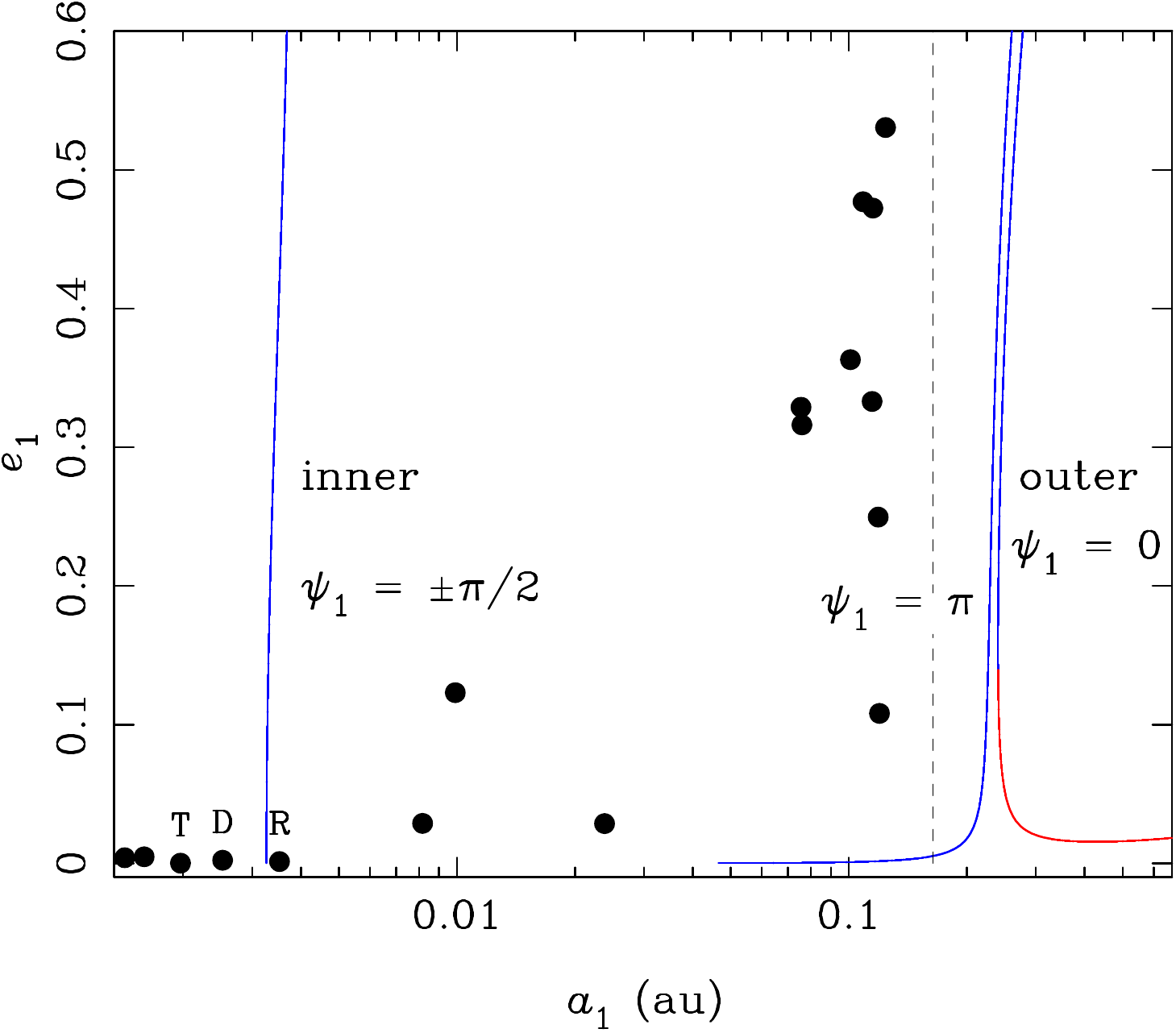}
\caption{Location of the inner and outer evections (full lines) for the masses and eccentricities of the Dione-Helene system. The stable branches (related to the libration centers) are shown in blue, and the unstable branch (related to the saddle point of the outer evection) is plotted in red. 
The full dots indicate the current position of the prograde
satellites of Saturn, and the vertical dashed line indicates the outermost limit of the irregular satellites. The letters specify the locations of Dione (D), Tethys (T), and Rhea (R), respectively. 
For the outer evection, the stable branch at $\psi_{1}=180^{\circ}$ and $e_{1} \protect\la 0.14$ is related to a kinematic libration (Fig. \protect\ref{topout}b).}
\label{evec-loc}
\end{figure}

\begin{figure*}
\centering{}
\includegraphics[width=0.29\textwidth]{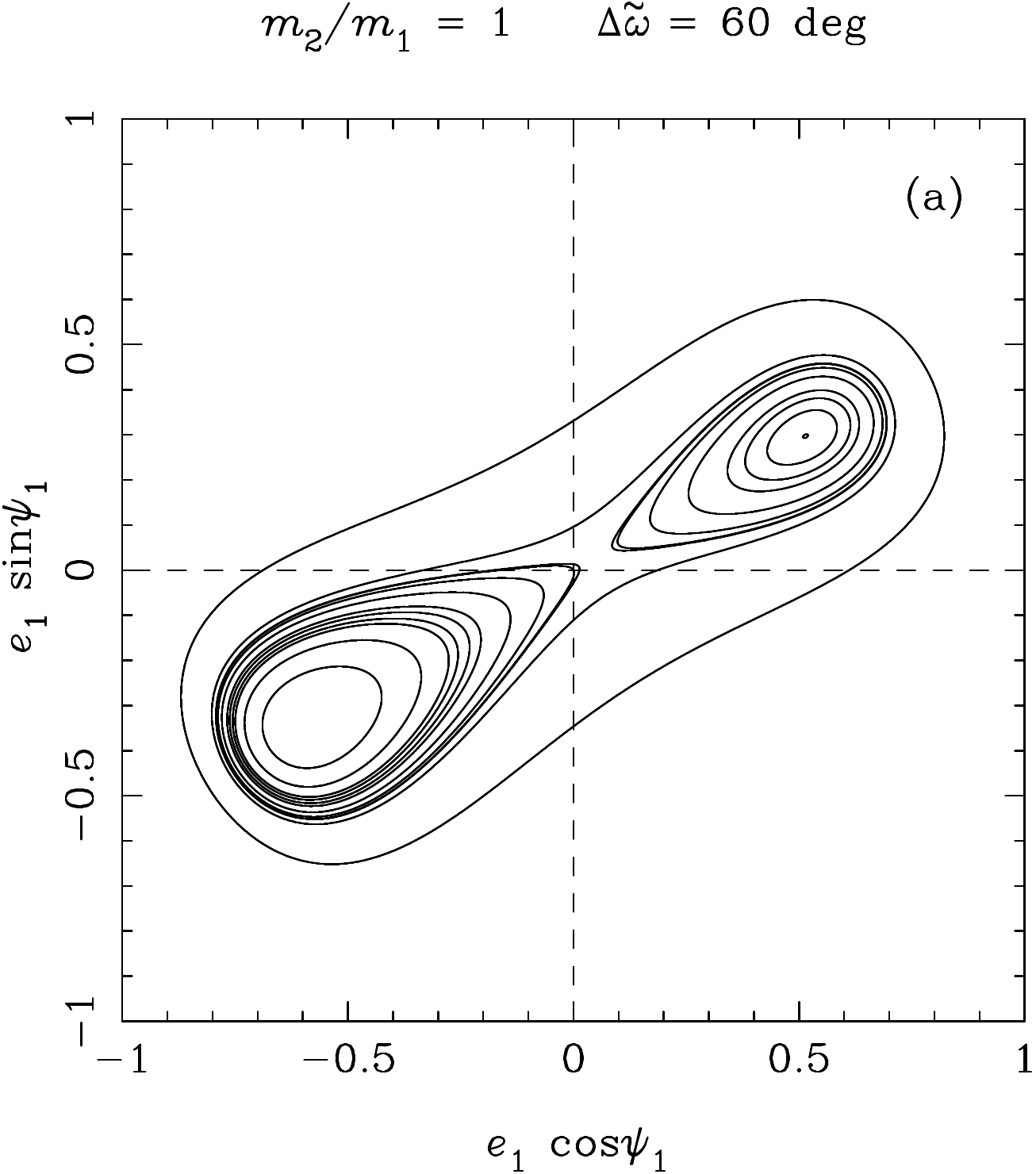}
\includegraphics[width=0.29\textwidth]{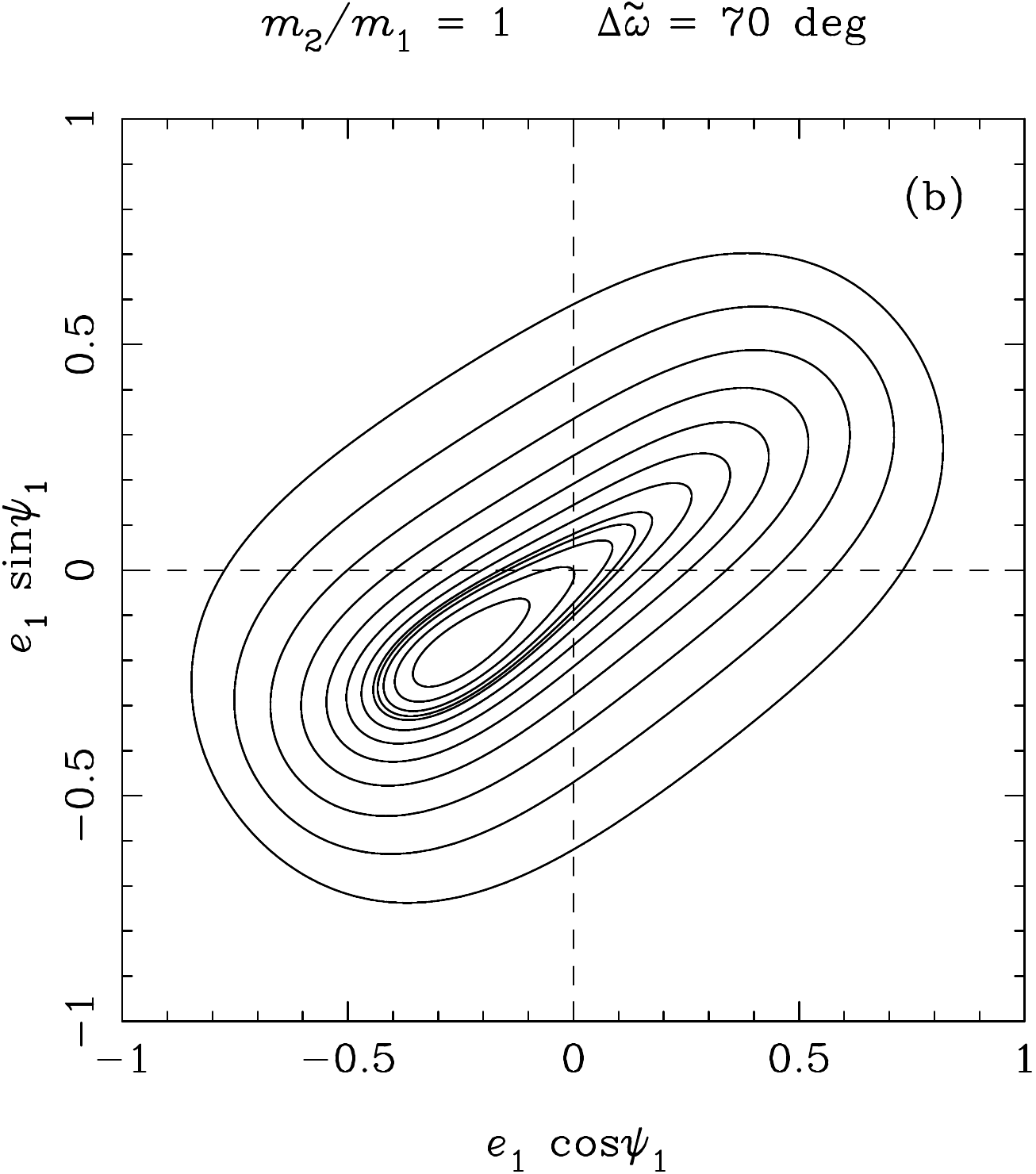}
\includegraphics[width=0.39\textwidth]{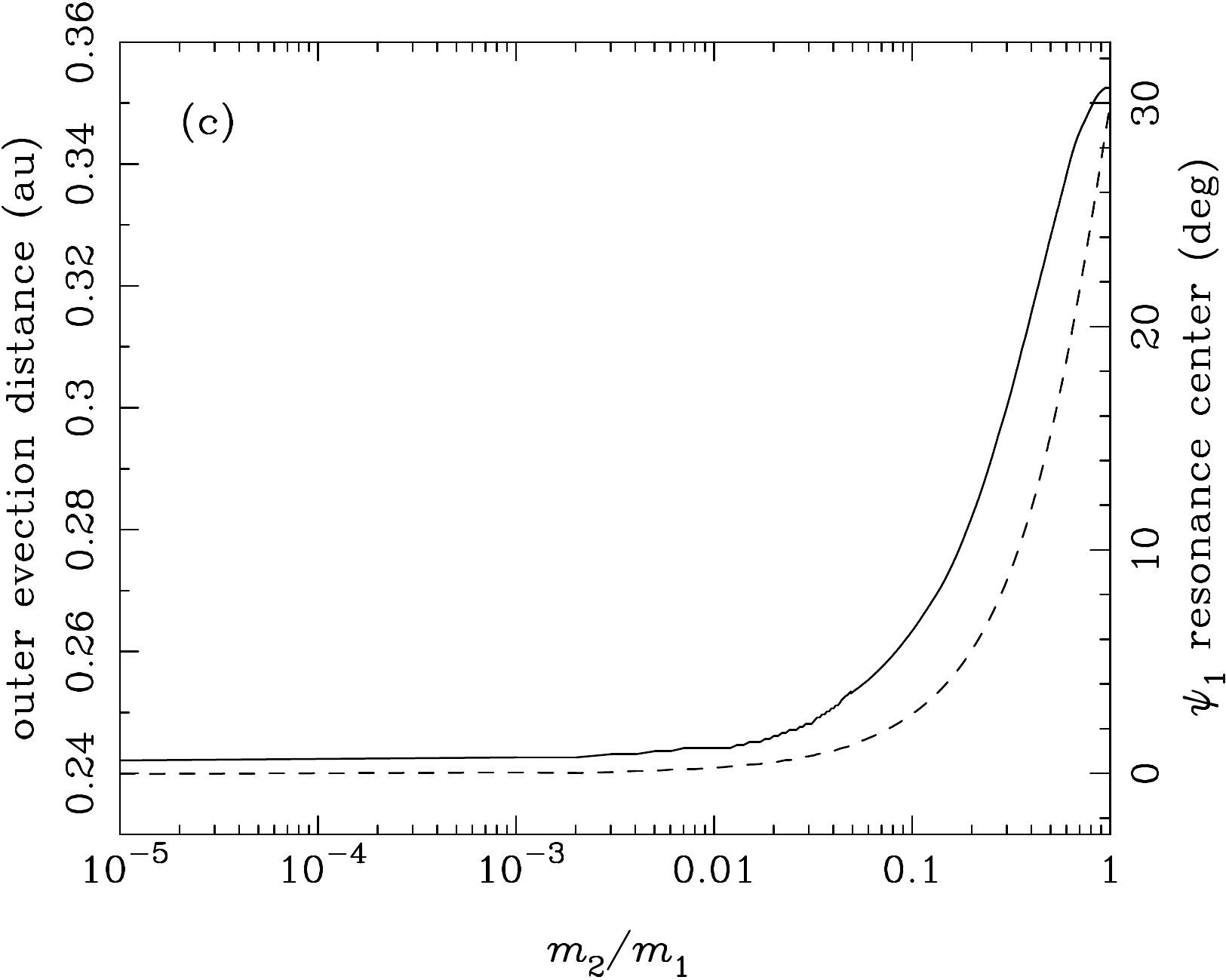}
\caption{Behavior of the outer evection resonance for equal masses of the trojan pair. Panel (a) corresponds to the topology of $m_{1}$ at a distance of $a_{1}=a_{2}=0.39$ au ($5.88\times10^{7}$ km), and assuming a periastron difference $\Delta\varpi =60^{\circ}$ (i.e., at the ACR solution).
Panel (b) is the same as (a), but for a periastron difference $\Delta\varpi =70^{\circ}$. Panel (c) shows the distance from the planet at which the evection resonance appears (full line), and the angle of the resonance center (dashed line) as a function of the mass ratio.}
\label{topout2}
\end{figure*}

The Hamiltonian Eq.~(\ref{eq:hevec}) represents a system with three degrees of
freedom that might be reduced in principle, provided that
the degrees of freedom evolve on very different timescales. This
would allow the application of the adiabatic invariance principle
to eliminate degrees of freedom, as in \citet{beauge_roig_2001} or \citet{Saillenfest_2016, Saillenfest_2017}. This rigorous semi-analytical treatment is beyond
the scope of this work. Here, we instead use a very simple zeroth-order
approach to provide an overview of the Hamiltonian topology. We know that in the absence of other perturbations, the stable equilibrium
of the trojan co-orbital motion arises when $\Delta\lambda =\lambda_{2}-\lambda_{1}\simeq 60^{\circ}$, $\Delta\varpi =\varpi_{2}-\varpi_{1}\simeq 60^{\circ}$, and $e_{2}\simeq e_{1}$ \citep{giuppone_etal_2010}. This corresponds to the so-called asymmetric apsidal corotation resonance solution, 
or ACR, which is schematically represented in Fig.~\ref{fig-orientation}. Therefore, we may freeze two degrees of freedom in Eq. (\ref{eq:hevec}) by fixing $\theta_{1},Z_{1}$
at one of the stable equilibrium points of $\bar{\mathcal{H}}_{0}$,
and also setting $\psi_{1}-\psi_{2}=60^{\circ}$ and $e_{1}=e_{2}$. In this way, the functions $A_{ki},B_{k},C_{k},\text{and }D_{i}$ become constant coefficients.
The resulting level curves of the remaining Hamiltonian of one degree of freedom are shown in Figs.~\ref{topoin}--\ref{topout2}.

It is worth recalling that in freezing the system at the ACR solution, we avoid the occurrence of other possible equilibria of the Hamiltonian with three degrees of freedom in Eq. (\ref{eq:hevec}). Moreover, we cannot be sure a priori that the ACR solution represents an equilibrium solution when the solar perturbation and the oblateness perturbations are taken into account. Nevertheless, the numerical experiments presented in Sects. \ref{sec.trojans} and \ref{tides} demonstrate that the ACR assumption is still valid in this case, and the ACR is an effective equilibrium state of the system. Exploring other equilibria would require searching for the zero-amplitude solutions given by the stationary conditions:
\bea
\label{eq_hamil}
\frac{\partial \bar{\mathcal{H}}}{\partial \Theta_i}= \frac{\partial \bar{\mathcal{H}}}{\partial J_i}=0 ,
\eea
where $J_i$ are the actions ($Z_1,W_1,W_2$) and $\Theta_i$ are the angles ($\theta_1,\psi_1,\psi_2$). These solutions represent the extrema of Eq. (\ref{eq:hevec}) in a six-dimensional space and could be assessed using numerical 
\citep[e.g.,][]{Giuppone_etal_2016}, semi-analytical \citep[e.g.,][]{giuppone_etal_2010}, or analytical \citep[e.g.,][]{Robutel_2013} techniques. Such an analysis, however, is beyond the scope of this work.

\subsection{Application to the Saturn system}
We applied our model to a physical system consisting of Saturn with $m_0=2.858 \times 10^{-4}\,M_\odot$, $R_0 = 60\,268$~km, $J_2 = 1.6298 \times 10^{-2}$, $J_4=0$, a distance to the Sun $a_3=9.537$~au, and assuming $e_3=0$ (we recall that the eccentricity of Saturn is 0.054). We considered satellite masses on the order of those of Tethys and Dione or lower ($\leq 6$-$10\times 10^{20}$~kg), and varied the mass ratio $m_2/m_1$ to describe the dynamics of the model. Hereafter, we refer to the angles $\psi_i$ as the evection angles.

Two different resonant regimes can be distinguished. The first occurs at very close distances from the central body and is driven
by the perturbation of the planet oblateness. This regime corresponds
to the inner evection resonance, and its topology displays two symmetric libration islands that resemble a second-order Andoyer Hamiltonian. For low mass ratios $m_{2}/m_{1}$, the libration center of the trailing trojan, $m_{1}$, is located at $\psi_{1}=\pm 90^{\circ}$ (Fig.~\ref{topoin}a). This topology is similar to the case of a single satellite (i.e., $m_{2}=0$). On the other hand, the behavior of the leading trojan $m_{2}$ is exactly the same, but the libration centers are shifted to $\pm 90^{\circ}-\Delta\varpi$ (Fig.~\ref{topoin}b). 
When $m_{2}=m_{1}$, the topology is preserved, but the libration centers are shifted to $\pm 90^{\circ} + \Delta\varpi /2$ for the trailing and to $\pm 90^{\circ} - \Delta\varpi /2$ for the leading trojan (Figs.~\ref{topoin}c,d). It is worth noting that the level curves in Fig.~\ref{topoin} were computed for the same semi-major axes (i.e., the same values of $Z_{1}$), implying that the distance at which the inner evection resonance occurs is roughly independent of the mass ratio.

The second resonant regime occurs at large distances from the planet and is driven by the solar perturbation. This regime corresponds to the outer (or classical) evection resonance. The basic topology shows two asymmetric libration islands enclosed by a separatrix generated by a single saddle point close to the origin (Fig.~\ref{topout}a). The asymmetry is caused by the presence in the disturbing function of the odd-degree terms in eccentricity, as previously shown by \cite{Frouard_2010}, in particular the term $e_i\cos\psi_i$, which is associated with the variation harmonic in the classical lunar theory.
For low $m_2/m_1$ ratios, the libration center of the trailing trojan is located at either $\psi_{1}=180^{\circ}$ or $0^{\circ}$, and this is again similar to the evection topology in the case of a single satellite. It is worth noting that for closer distances to the planet, the topology is not strictly resonant because there is no separatrix (Fig.~\ref{topout}b). In this case, we observe a kinematic libration around $\psi_{1}=180^{\circ}$ forced by the odd-degree terms in $e$.
On the other hand, since the unstable equilibrium that generates the separatrix is always shifted from the origin, a resonant orbit around $\psi_{1}=180^{\circ}$ but very close to the separatrix may {eventually} display a circulating angle $\psi_{1}$. Figure~\ref{topout}c shows the topology of the leading trojan, which is the same as in Fig.~\ref{topout}a, but rotated by $-\Delta\varpi$.

We recall that in the case of a single satellite, the precession rate caused by the oblateness of Saturn is $\dot \varpi_\mathrm{obl} \approx 3 n_1 J_2 R_0^2/(2a_1^2)$ \citep{Roy1978}, while the precession rate caused by solar perturbation is $\dot \varpi_\mathrm{sol} \approx 3 G m_3/(4 n_1 a_3^3)$ \citep{Innanen1997}.
Then, according to  \citet{Li_etal2016}, the critical semi-major axis at which one precession rate dominates  is
\begin{equation}
a_\mathrm{crit}=\left(2\frac{m_0}{m_3}J_2 R_0^2a_3^3\right)^{1/5}\ \sim 0.0167\,\mathrm{au}
,\end{equation}
which defines the transition limit between the inner and outer evections in the Saturn system.

The locations of the libration centers of the inner and outer evection resonances,
computed for the parameters of the Dione-Helene system ($m_{2}/m_{1} \simeq 2\times 10^{-5}$), are shown
in Fig.~\ref{evec-loc}. 
These locations were obtained by searching for the conditions in the $a,e$ plane that make 
$\partial \bar{\mathcal{H}}/\partial W_{1}=0$. The outer evection shows two branches, one corresponding to $\psi_{1}=180^{\circ}$ , which is always stable (blue line),
and the other corresponding to $\psi_{1}=0^{\circ}$ , which splits
into the stable (blue line) and the unstable (red line) equilibria.

Figure~\ref{evec-loc} also shows the position of the prograde moons of Saturn. We note that all the moons are located inside of the location of the outer evection resonance, including the group of retrograde moons (not shown), whose outermost limit is indicated by the vertical dashed line.
We also note that the regular moons, that is, those with $e \la 0.02$ and $i \la 1^\circ$, are clustered inside of the location of the inner evection resonance ($a\la 0.003$~au). In particular, all the moons involved in co-orbital configurations (Janus-Epimetheus, Tethys, and Dione) are in this group. The only exception among the regular moons is Rhea, which is located slightly outside of the inner evection, but has no co-orbitals.

In the case of equal masses, the topology of the outer evection resonance is preserved, but the libration centers of both the trailing and leading trojans are rotated by $\pm \Delta\varpi /2$, respectively (e.g., Fig.~\ref{topout2}a). At variance with the inner evection resonance, the existence of the libration islands in this case is very sensitive to the value of $\Delta\varpi$, and they quickly disappear if the trojan pair moves away from the ACR solution toward increasing $\Delta\varpi$ (Fig.~\ref{topout2}b). Another major difference with respect to the inner evection is that the distance from the planet at which the resonant islands appear (more precisely, where the unstable equilibrium appears) is strongly dependent on the mass ratio, as shown in Fig.~\ref{topout2}c (full line). This figure also shows how the libration center moves away from $\psi_{1}=0$ as the mass ratio tends to 1 (dashed line).

\section{Numerical study}\label{sec.methods}

\begin{figure*}
\centering                                     
\includegraphics[width=0.32\textwidth]{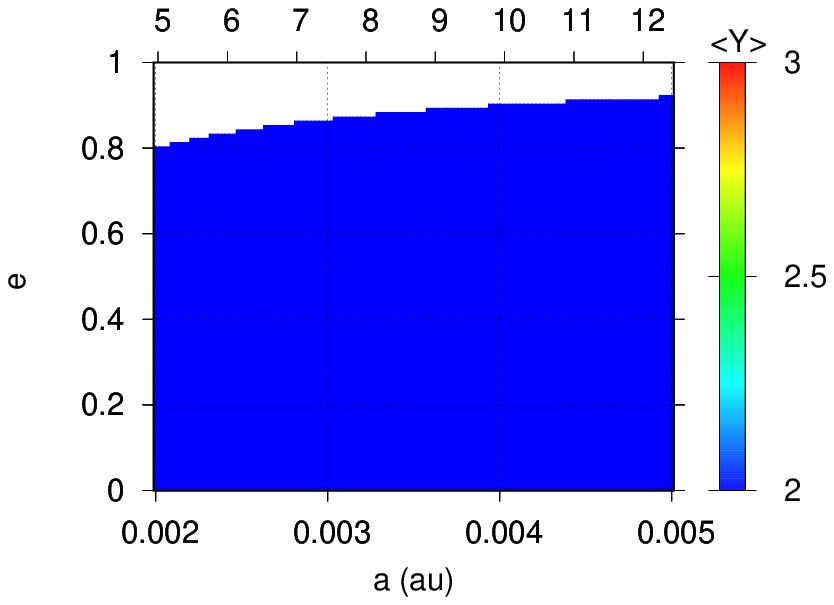}  \includegraphics[width=0.32\textwidth]{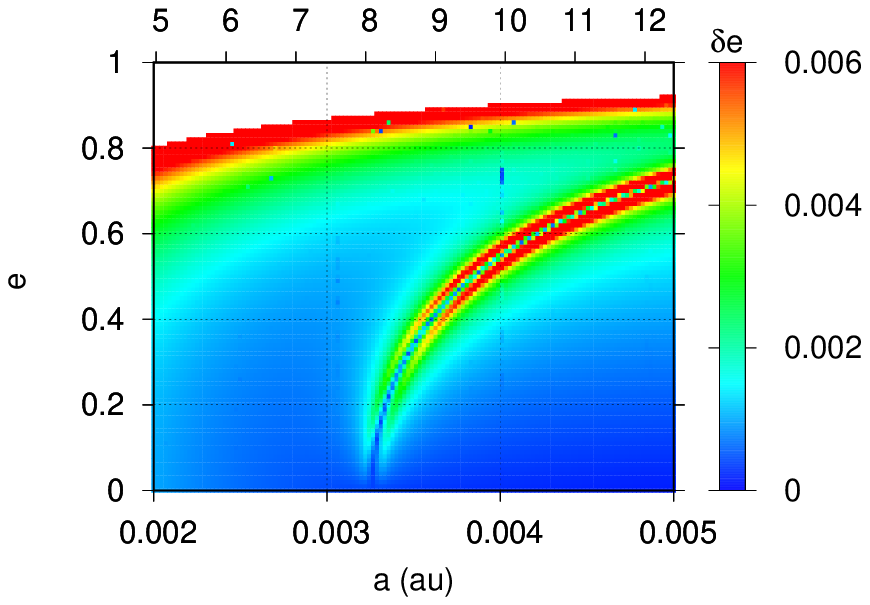} 
\includegraphics[width=0.32\textwidth]{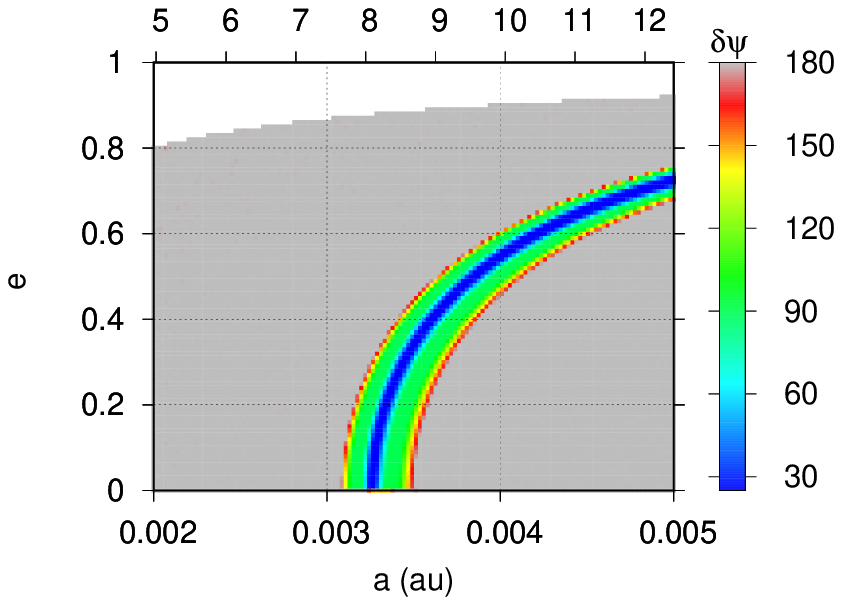}  \\
\includegraphics[width=0.32\textwidth]{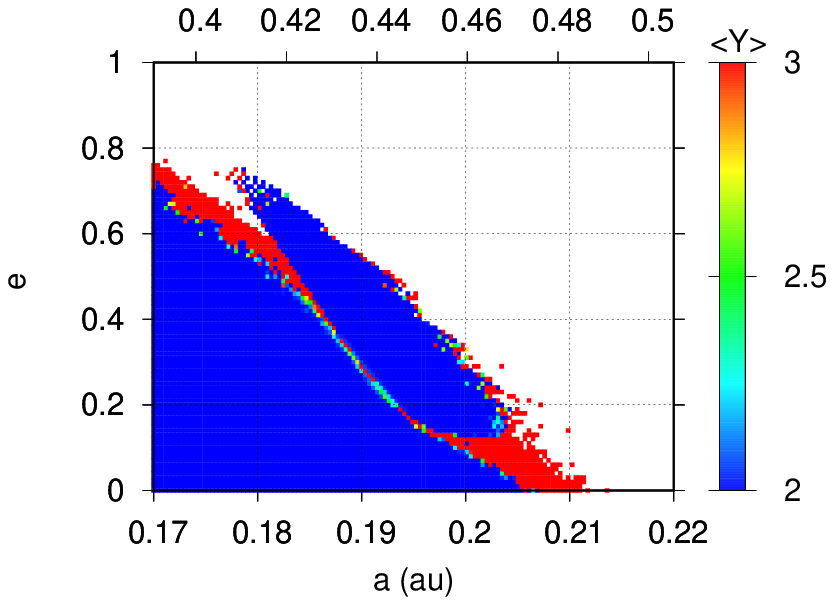}  \includegraphics[width=0.32\textwidth]{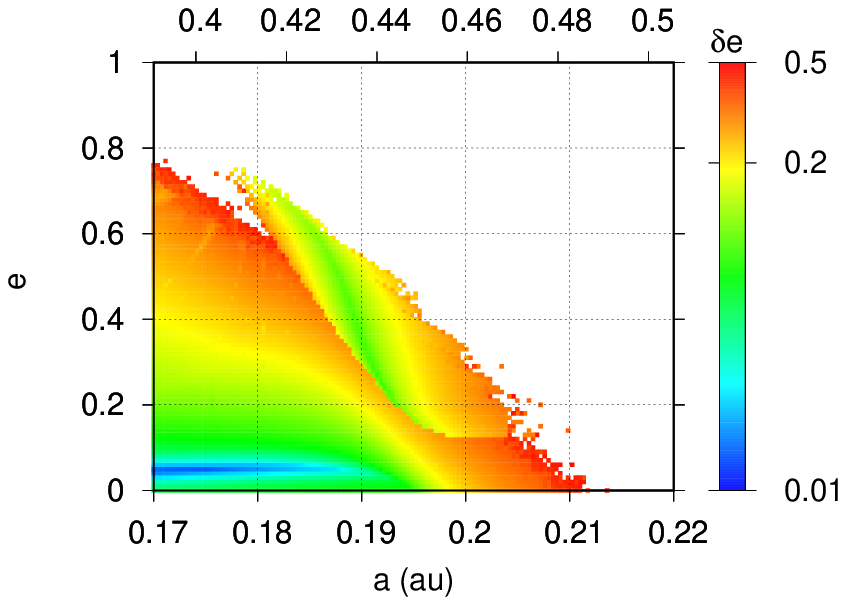} \includegraphics[width=0.32\textwidth]{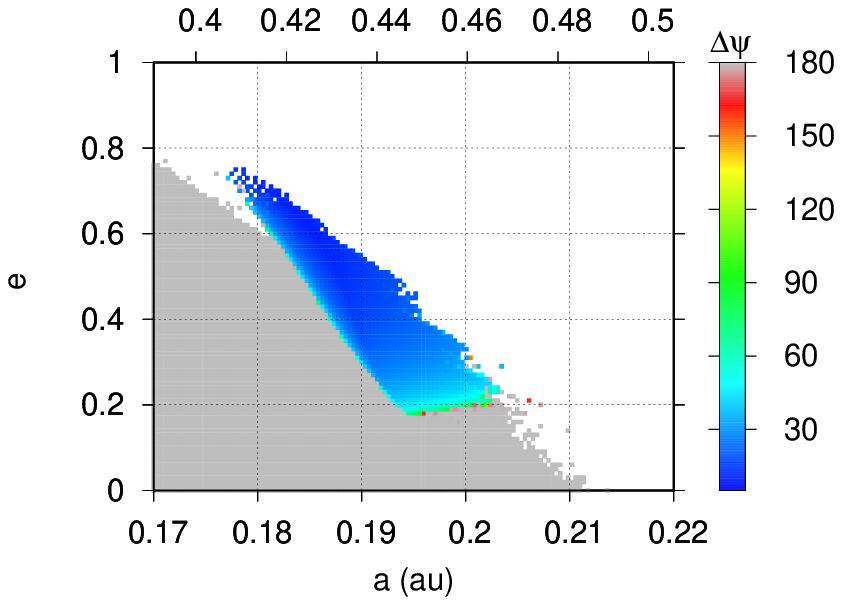} 
\caption{Dynamical maps in the $(a,e)$-plane for the regions of the inner evection (top row) and the outer evection (bottom row). The MEGNO, $\delta e$, and $\delta\psi$ indicators are displayed in the left, middle, and right columns, respectively. Angles are in degrees. White areas correspond to unstable orbits. The gray areas in the $\delta\psi$ maps indicate non-librating orbits. The grids correspond to osculating orbital elements with initial $\psi =90^{\circ}$ for the inner evection, and initial $\psi =0^{\circ}$ for the outer evection. To allow comparison with previous works, the upper horizontal axes give the distance to the planet in convenient units: planetary radii ($R_\mathrm{S}$) for the inner evection, and Hill radii ($R_\mathrm{Hill}$) for the outer evection.}
\label{fig-Frouardzoom}
\end{figure*}
  
In this section we present a numerical survey of the stability of the Saturn satellites, with particular focus on the occurrence of the evection resonance in both the cases of a single satellite and a trojan pair. Our numerical model takes into account the perturbations from the Sun as well as from the oblateness of Saturn by including the $J_2$ zonal {harmonic}. The perturbations from other satellites in the Saturn system are not taken into account.
The satellite orbits are described by the orbital elements $a,e,\lambda,\text{and }\varpi$. Hereafter, we maintain the subscripts 1,2 for the trojan pair (trailing and leading, respectively), and we use the subscripts $\odot$ and S for the Sun and Saturn, respectively.

The Newton equations of motion are numerically solved using a Burlisch-Stoer integrator with adaptive step size, which was modified to independently monitor the error in each variable; this imposes a relative precision better than $10^{-13}$. We stopped the integrations when the mutual distance between Saturn and the satellite was smaller than the sum of their radii. The total integration time was set to encompass several periods of the orbital secular variations.

The stability was evaluated through the use of some chaos and resonance indicators, and we explored different grids of initial conditions in the space of orbital elements. In particular, for each initial condition of the grid, we computed the value of the indicator for the mean exponential growth of nearby orbits (MEGNO), $\left\langle Y\right\rangle_i$, $i=1,2$ \citep{Cincotta_2000, Maffione_etal_2011}. This indicator is very efficient to quickly distinguish regular ($\left\langle Y\right\rangle_i \sim 2$) from chaotic orbits ($\left\langle Y\right\rangle_i \gg 2$), but it does not provide a detailed representation of the structure of a resonance. 

To better address the structure of the resonances, we also computed the amplitude of maximum variation of the orbital eccentricity of the satellites during the integrations, that is, $\delta e_i= \left(e_{i,\mathrm{max}}-e_{i,\mathrm{min}}\right)/2$, $i=1,2$.
This indicator has proven to be an extremely useful tool for mapping the resonant structure in $N$-body problems \citep[e.g.,][]{Ramos_2015}. We recall that $\delta e_i$ is not a direct measure of chaotic motion, but abrupt changes in $\delta e_i$ are often tracers for the presence of resonances. Therefore, regions with large variations in $\delta e_i$ are more sensitive to perturbations and are very likely chaotic. \citet{Frouard_2010} used a similar indicator in their study of the evection resonance in Jupiter satellites, defined as $(e_\mathrm{max}^{\tau_a} - e_\mathrm{max}^{\tau_b})/e_\mathrm{max}^{\tau_a}$, where the total integration time, $\tau$, was divided into two consecutive samples $\tau_a$ and $\tau_b$. Nevertheless, we realized that this indicator is quite sensitive to the value of $\tau$ and may artificially create some structures that are not related with actual resonances.

A third indicator that we used is the amplitude of oscillation of the evection angle, defined as $\delta\psi_i =\left|\psi_{i,\mathrm{max}}-\psi_{i,\mathrm{min}}\right|$. This indicator has a minimum at the center of the resonance, or more generally, at the condition $\dot{\psi}_i=\dot{\lambda}_{\odot}-\dot{\varpi}_i\sim 0$.

\subsection{Evection for a single moon}\label{sec.single}

\begin{figure}
\centering
\includegraphics[width=1\columnwidth]{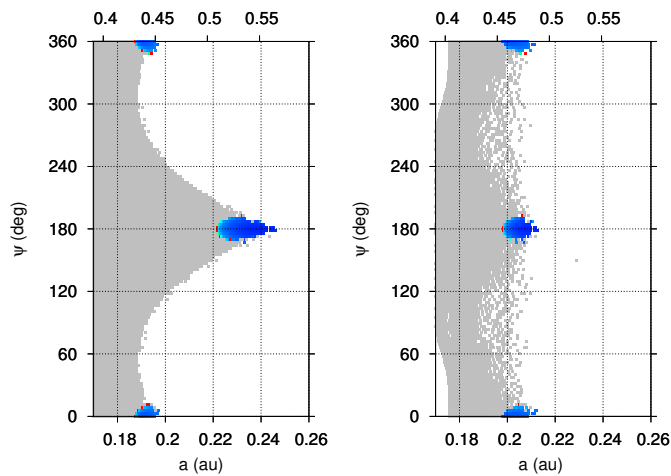}
\caption{Dynamical maps of the $\delta\psi$ indicator in the $(a,\psi)$-plane for the outer evection, with initial $e=0.4$. The left frame corresponds to a grid of osculating elements, while the right frame is the same grid in terms of mean elements. Blue regions represent the conditions for which $\delta\psi \protect\la 30^{\circ}$. Gray regions correspond to larger libration amplitudes or circulation of $\psi$. White regions are unstable orbits. The resonance occurs at a mean semi-major axis $\bar{a}=0.205$~au. The upper horizontal axis is the distance in terms of $R_\mathrm{Hill}$.}
\label{fig-incs00}
\end{figure}

\begin{figure}
\centering                                     
\includegraphics[width=0.9\columnwidth]{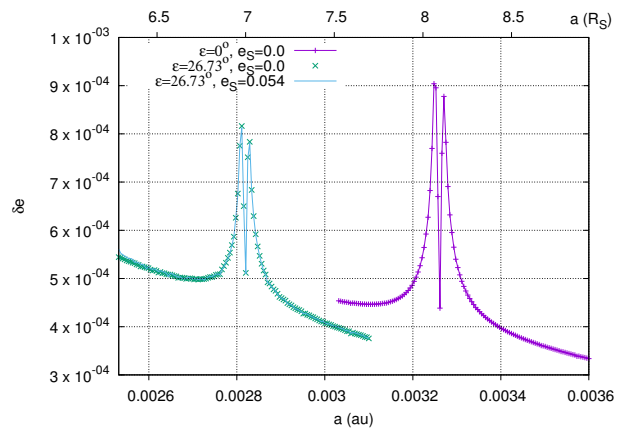}
\caption{Value of $\delta e$ as a function of the distance from the planet in the region of the inner evection and for $e=0$. The curve in violet corresponds to the case of an obliquity $\epsilon_\mathrm{S}=0^\circ$ and orbital eccentricity $e_\mathrm{S}=0^\circ$ for Saturn. The green dots correspond to the case of $\epsilon_\mathrm{S}=26.73^\circ$, either with $e_\mathrm{S}=0^\circ$ or $e_\mathrm{S}=0.054^\circ$. The minimum inside the peak of the curves marks the center of the evection resonance.} 
\label{fig-Jinn}
\end{figure}

\begin{figure}
\centering                       
\includegraphics[width=0.8\columnwidth]{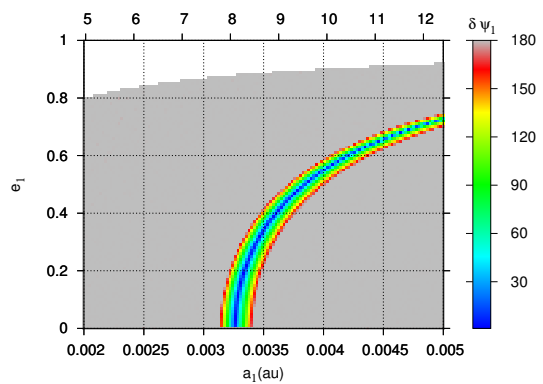} \\
\caption{Dynamical map of the $\delta\psi_1$ indicator in the osculating $(a,e)$-plane for the inner evection, setting $\psi_1=120^\circ$ initially. Case of trojan satellites with equal masses. The resonance occurs at the same location as in the case of a single satellite (compare to Fig.~\protect\ref{fig-Frouardzoom}). The upper horizontal axis gives the distance in $R_\mathrm{S}$. Angles are in degrees. }
\label{fig-maptrojans-int}

\end{figure}
  
\begin{figure*}[!ht]
\centering                                     
\includegraphics[width=0.8\textwidth]{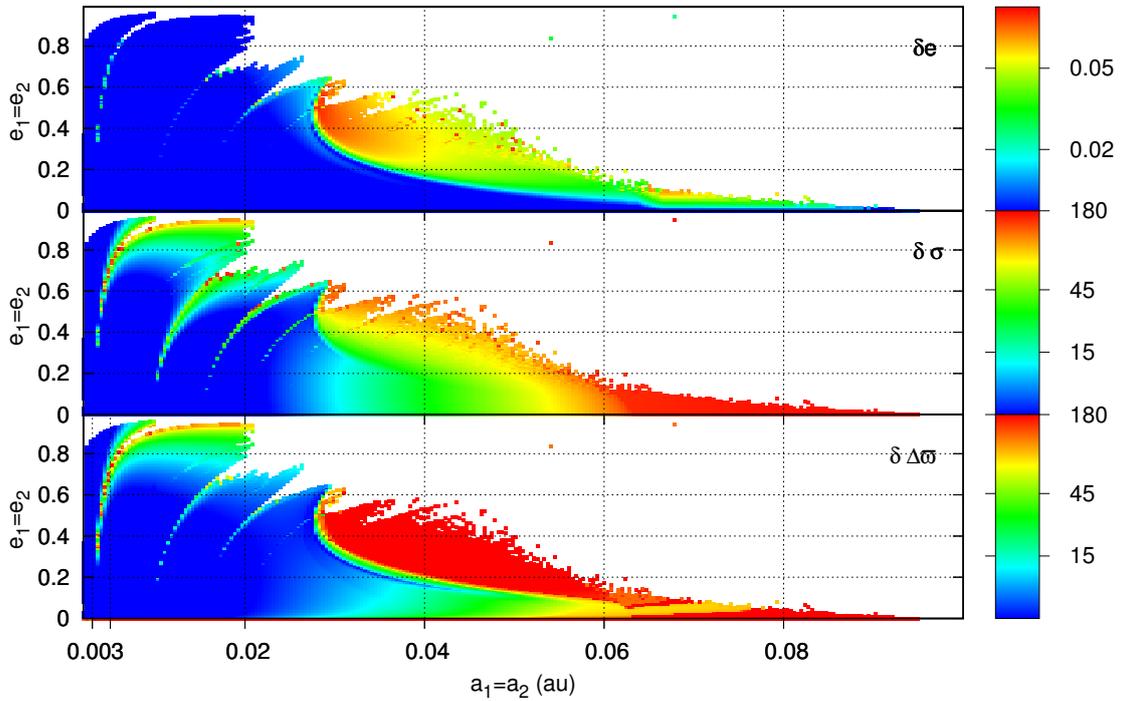}
\caption{Dynamical maps in the osculating $(a,e)$-plane displaying the amplitude $\delta e_1$ of the eccentricity variation (top), the amplitude of the $\sigma$ libration (middle), and the amplitude of the $\Delta\varpi$ libration (bottom) for a system of two equal-mass moons initially in a trojan ACR configuration. Angles are in degrees.
Trojan motion is unstable in the white regions in less than $5 \times 10^4$ years, and it is most stable in the blue regions. The stripes of less stable trojan motion observed for $a<0.03$~au correspond to the locations of the evection resonance and its higher harmonics.}

\label{fig-anglesp}

\end{figure*}

Figure~\ref{fig-Frouardzoom} shows a set of dynamical maps in the $(a,e)$-plane for the inner (top row, $\psi =90^{\circ}$) and the outer (bottom row, $\psi =0^{\circ}$) evections. The maps display the values of the three indicators described in Sect.~\ref{sec.methods}: $\left\langle Y\right\rangle$ (left column), $\delta e$ (middle column), and $\delta \psi$ (right column). We drop here the subindex $i$ for simplicity.

The inner evection region is quite regular, and both $\delta e$ and $\delta \psi$ allow identifying the exact location of the resonance. For eccentricities lower than $\sim 0.6$, this location coincides with the eccentricity predicted by the analytical model (cf. Fig.~\ref{evec-loc}). At higher eccentricities, the resonance location moves toward slightly larger semi-major axes, a behavior that the analytical model does not reproduce because the $\bar{\mathcal{J}}_i$ series is truncated early. The gray areas in the map of $\delta\psi$ correspond to conditions where the angle $\psi$ circulates. White areas correspond to orbits that either collide with Saturn ($a\leq R_\mathrm{S}$) or escape from the system ($a>a_\mathrm{S}$) during the integration time-span.

The outer evection resonance, on the other hand, occurs inside a stable island surrounded by a thin layer of chaotic motion (red in the map of $\left\langle Y\right\rangle$). The map of $\delta\psi$ shows that the libration amplitude around $\psi =0^{\circ}$ is never larger than $\sim 60^{\circ}$, in good agreement with the analytical model (cf. Fig.~\ref{topout}a). The map of $\delta e$ also shows a stable narrow region at low eccentricities (blue) that is related to the kinematic librations induced by the forced eccentricity mode. Nevertheless, we note that the resonant island occurs closer to the planet ($a\sim 0.19$~au) than predicted by the analytical model ($a\sim 0.26$~au). This is due in part to the fact that the grids in Fig.~\ref{fig-Frouardzoom} are given in terms of the osculating initial
semi-major axis, while in terms of the mean semi-major axis, the island shifts to higher values, as shown in Fig.~\ref{fig-incs00}. Another reason might be the early truncation of the disturbing function Eq.~(\ref{eq:hamt2c}) and that the eccentricity of the Sun is not considered in the model. A similar behavior has been identified by \citet{Frouard_2010}.

Figure~\ref{fig-incs00} shows the dynamical maps of $\delta\psi$ in the ($a,\psi$)-plane for the outer evection. In this case, we have initially $e=0.4$ and $\lambda_\odot=0^{\circ}$, which means that varying the initial $\psi$ is equivalent to varying the initial $\varpi$. We note that the libration regions around $0^\circ$ or $180^\circ$ occur at different osculating semi-major axes (Fig.~\ref{fig-incs00}, \textit{left}), but this dependence disappears when the mean semimajor axis is considered (Fig.~\ref{fig-incs00}, \textit{right}).

Since the inner evection is driven by the planet oblateness, it is expected that the planet obliquity plays some role in the behavior of the resonance. In Fig.~\ref{fig-Jinn} we address the effect of the obliquity of Saturn ($\epsilon_\mathrm{S} =26.73^\circ$) and its orbital eccentricity ($e_\mathrm{S} =0.054$) on the inner evection. We note that a non-zero obliquity shifts the location of the resonance to slightly smaller distances (from $\sim 8$ to $\sim 7\,R_\mathrm{S}$), although the structure of the equilibrium family is preserved. On the other hand, the eccentricity of the Saturn orbit (or equivalently, the Sun) does not produce any significant changes.

\subsection{Evection for trojan moons}\label{sec.trojans}
  
In this section, we study the evolution of two co-orbital moons in trojan configuration initially near the equilibrium of the ACR condition, that is, $\sigma \equiv \Delta\lambda=60^\circ$, $\Delta\varpi=60^\circ$, $e_2=e_1$ \citep{giuppone_etal_2010, Robutel_2013}.  We consider two cases: $m_2/m_2=1$ and $m_2/m_1 =10^{-4}$. All the simulations initially have $\lambda_\odot=0^{\circ}$ . 
Figure~\ref{fig-orientation} shows a schematic orbital diagram when $\psi_2=0^\circ$.

Figure~\ref{fig-maptrojans-int} shows the dynamical map of the $\delta \psi_i$ indicator for a system of two trojan moons with equal masses in the region of the inner evection. The indicator corresponds to those of the trailing trojan $(i=1)$, but the situation is identical for the leading trojan. A comparison to Fig.~\ref{fig-Frouardzoom} shows that the evection resonance occurs at almost the same distance from the planet as for a single moon ($a\sim0.003$~au at $e\sim0$), implying that the location of the inner evection is independent of the mass ratio of the trojans. This agrees well with the predictions of the analytical model (Fig.~\ref{topoin}).

On the other hand, the outer evection in the case of trojan moons cannot be studied because the trojan configuration does not survive at large distances from Saturn. This is shown in Fig.~\ref{fig-anglesp} for moons of equal masses. The white regions correspond to the conditions where an initial trojan ACR configuration is destroyed, and the co-orbital pair does not survive as such. Stable trojan configurations exist up to distances of $\sim0.06$~au only for almost circular orbits. At distances $\protect\la 0.02$~au (blue regions), the trojan ACR configuration is stable even for high eccentricities, except at the locations of the evection resonance and its higher harmonics ($\dot{\varpi}_i\sim\dot{\lambda}_{\odot}$; $\dot{\varpi}_i\sim 2\dot{\lambda}_{\odot}$; $\dot{\varpi}_i\sim 3\dot{\lambda}_{\odot}$). The same qualitative results are obtained for a trojan pair with $m_2/m_1 = 10^{-4}$.

\begin{figure}
\centering                                     
\includegraphics[width=1\columnwidth]{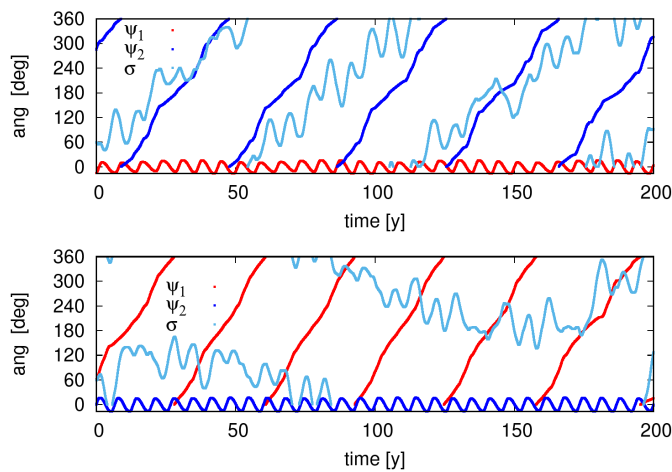}
\caption{Evolution of the resonant angles $\sigma$ (cyan lines) and $\psi_i$ (red/blue lines) for a pair of moons with equal masses, {initially placed} in a trojan ACR configuration ($\sigma=60^\circ$, $\Delta\varpi=60^\circ$) in the region of the outer evection ($a_1=a_2 \sim 0.2$~au, $e_1=e_2=0.4$). The top panel corresponds to the case $\psi_1=0^\circ$ and $a_1=a_2=0.1825$~au initially. The bottom panel corresponds to the case $\psi_2=0^\circ$ and $a_1=a_2=0.1893$~au initially.} 
\label{fig-evol}
\end{figure}

\begin{figure}
\centering
\includegraphics[width=1\columnwidth]{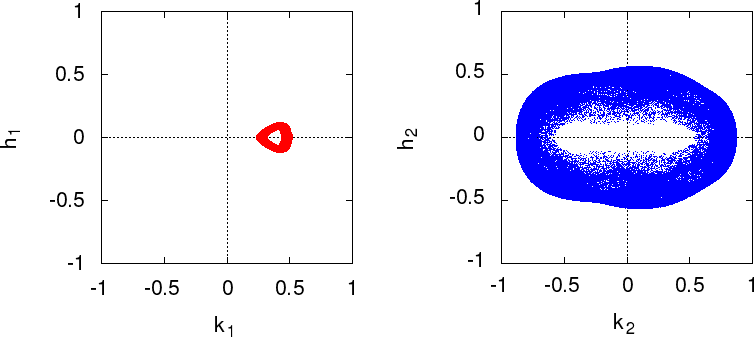} \\
\includegraphics[width=1\columnwidth]{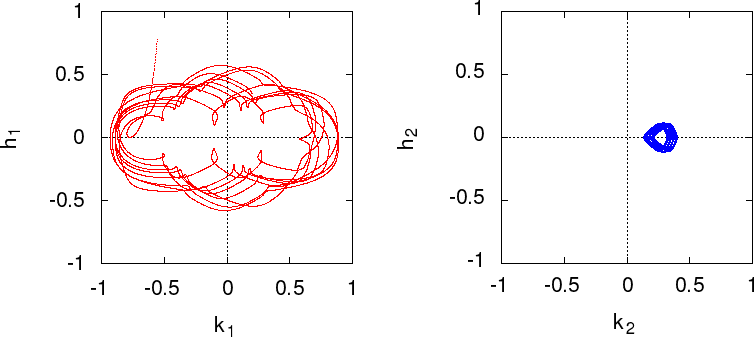} 
\caption{{Same as Fig.~\protect\ref{fig-evol}, but in the $k_i,h_i$ space (see text) and for the full time span of the simulation. The top panels correspond to the top panel of Fig.~\protect\ref{fig-evol}; the system survives for at least $10^6$~yr. The bottom panels correspond to the bottom panel of Fig.~\protect\ref{fig-evol}; the system survives for $\sim 10^3$~yr. In both cases, $\lambda_\odot=0^\circ$ initially.} }
\label{fig-kh1}
\end{figure}

In the region where the outer evection resonance occurs for single moons ($a\sim0.2$~au, see Fig.~\ref{fig-Frouardzoom}), trojan configurations are not sustainable for more than 100~yr. As a consequence, a pair of moons initially inside the evection island and in a trojan ACR configuration quickly ends up as a non-trojan pair with only one of the components trapped in the evection resonance. {This is shown in Fig.~\ref{fig-evol} for a pair of equal-mass moons. The top panel shows the case where $\psi_1$ librates while $\psi_2$ circulates, and the bottom panel shows the opposite case. Neither of these cases show $\sigma$ librating around $\pm 60^\circ$, so trojan motion does no longer exist. Simultaneous libration of $\psi_1,\psi_2$ is never found in the outer evection either. Figure~\ref{fig-kh1} shows the same evolutions as Fig.~\ref{fig-evol}, but in the space $k_i = e_i\cos\psi_i,h_i = e_i\sin\psi_i$.}

\begin{figure}
\centering
\includegraphics[width=1\columnwidth]{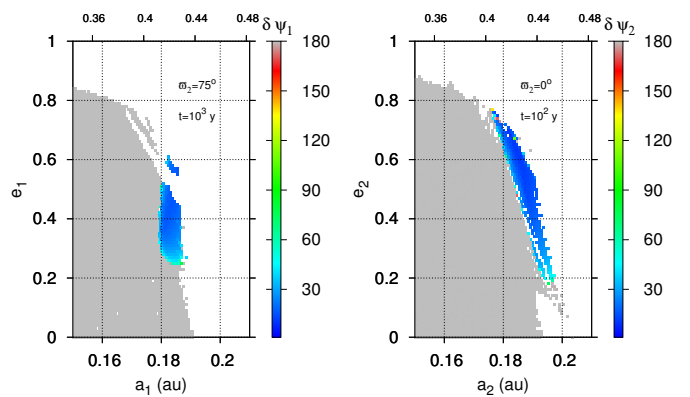}
\caption{Dynamical maps of $\delta\psi_i$ in the osculating ($a,e$)-plane for a trojan pair of equal masses, {initially placed} in the ACR configuration. Initial conditions for the angles are the same as in Fig.~\protect\ref{fig-kh1}.
Blue islands correspond to libration of $\psi_1$ (left) or $\psi_2$ (right). Gray regions correspond to circulation of $\psi_i$ , and white regions correspond to unstable orbits (compare to Fig.~\protect\ref{fig-Frouardzoom}). The upper horizontal axis is the distance in terms of $R_\mathrm{Hill}$.}
\label{fig-maptrojans-ext}
\end{figure}

The libration of either $\psi_1$ or $\psi_2$ is related to the initial longitudes of the periastrons relative to the longitude of the Sun (e.g., Fig.~\ref{fig-orientation}). The behavior of the former trojan companion that becomes capture in the evection resonance is very similar to that of a single moon. In the ($a,e$)-plane, the resonance occurs inside an island surrounded by a layer of chaotic motion, as shown in Fig.~\ref{fig-maptrojans-ext}. We note, however, that the island corresponding to the libration of $\psi_1$ (left panel) is smaller than that corresponding to the libration of $\psi_2$ (right panel), indicating that even in the case of equal masses, the configuration may not be symmetric. Neither configuration is long-term stable, and at some point, the moon where $\psi_i$ circulates is ejected or hits the other moon or the planet. Nevertheless, configurations that lead to the libration of $\psi_1$ appear to be more stable (up to $10^6$~yr) than those leading to the libration of $\psi_2$ (up to $\sim 10^3$~yr; see Fig.~\ref{fig-kh1}).
        
\section{Tidal evolution}\label{tides}

\begin{figure*}
\centering
\includegraphics[width=0.32\textwidth]{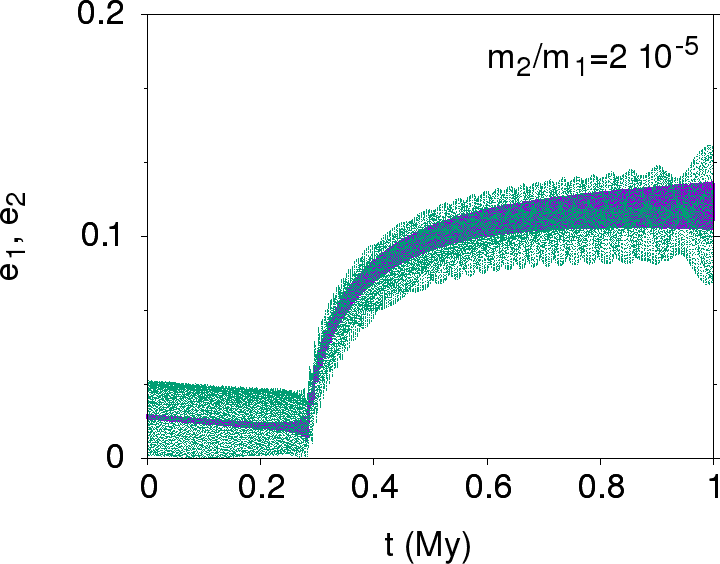} 
\includegraphics[width=0.32\textwidth]{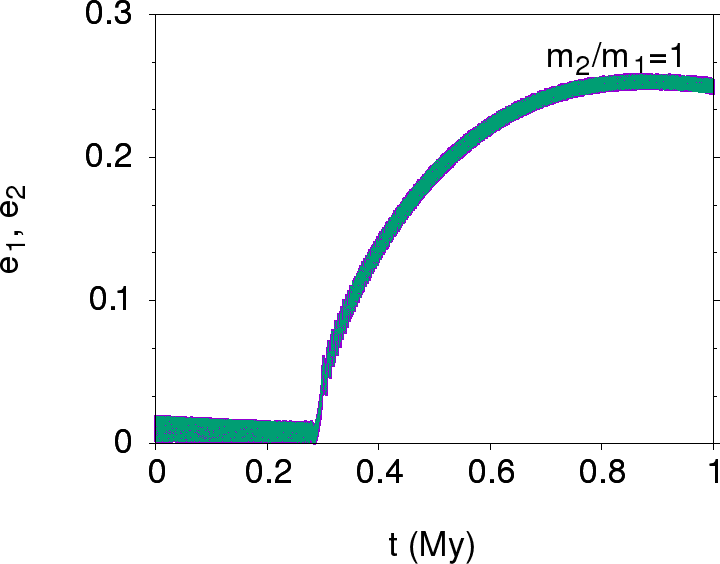} 
\includegraphics[width=0.32\textwidth]{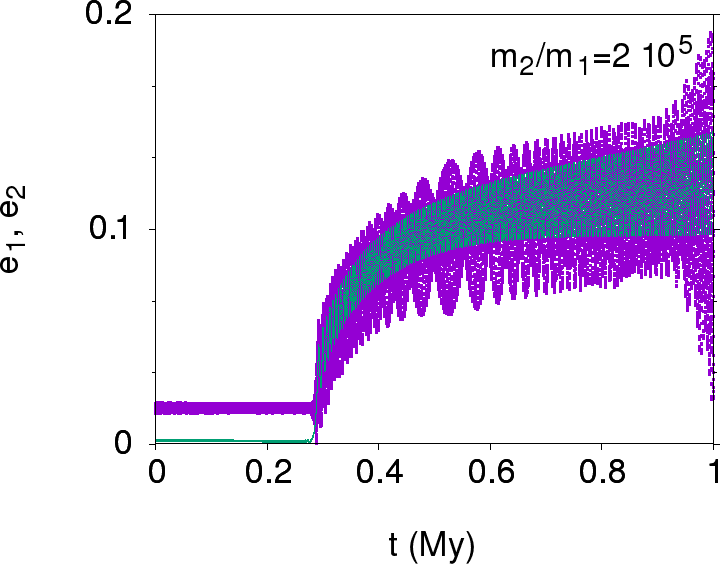} \\
\includegraphics[width=0.32\textwidth]{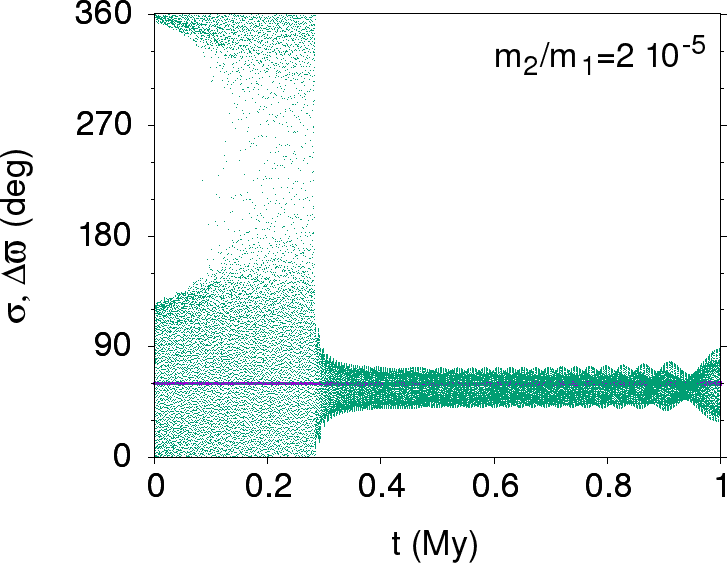}   
\includegraphics[width=0.32\textwidth]{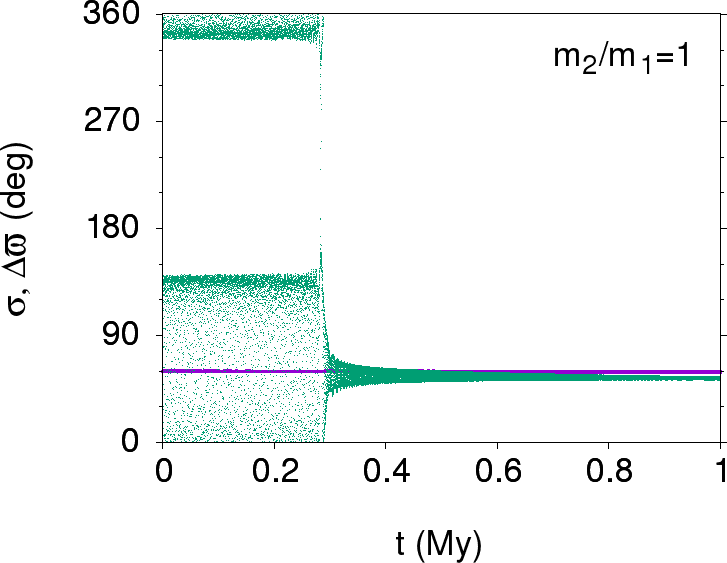} 
\includegraphics[width=0.32\textwidth]{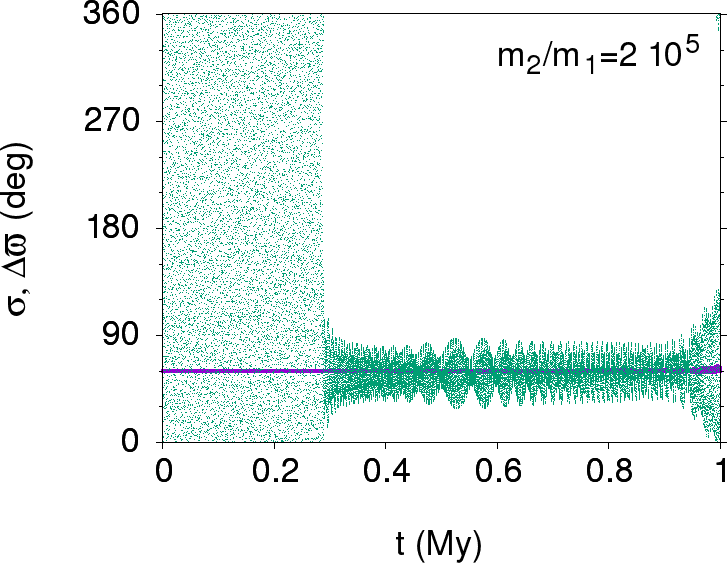}   \\
\includegraphics[width=0.32\textwidth]{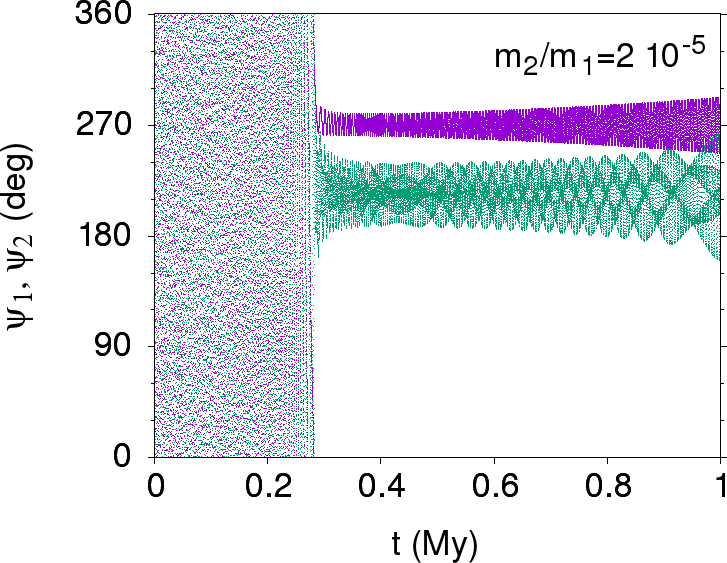} 
\includegraphics[width=0.32\textwidth]{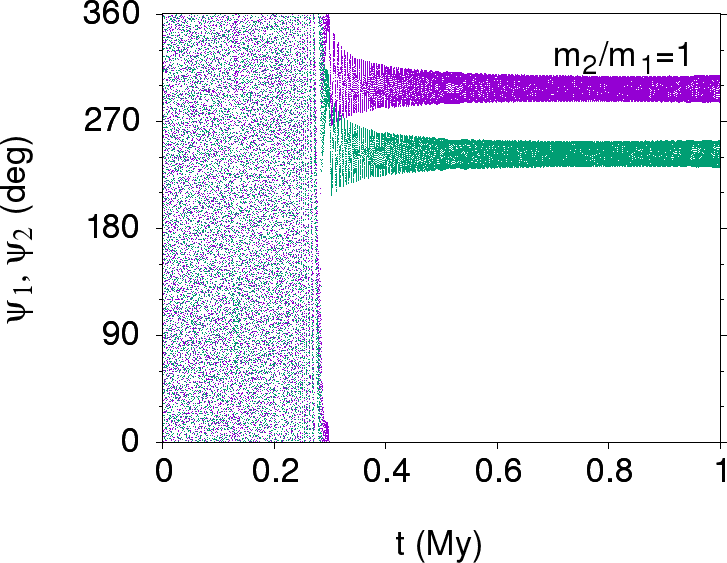} 
\includegraphics[width=0.32\textwidth]{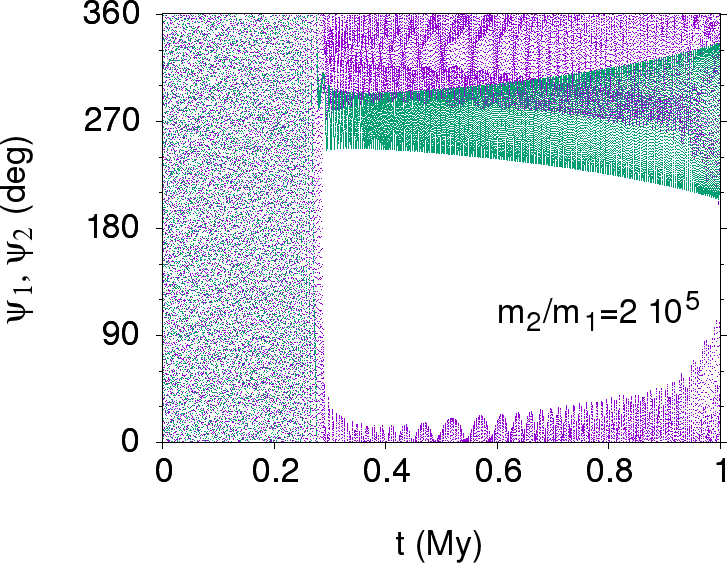} \\
\caption{Adiabatic evolution of pair of trojan satellites that migrate as a result of tidal effects, using $Q=100$. Each column of panels corresponds to a different mass ratio $m_2/m_1$: $2\times10^{-5}$ (left), 1 (middle), and $2\times10^{5}$ (right). Eccentricity is shown in the top row, $\sigma$ and $\Delta\varpi$ in the middle row, and $\psi_1,\psi_2$ in the bottom row. Magenta and green identify the trailing and leading trojan, respectively, except in the middle panels, where they identify $\sigma$ and $\Delta\varpi$, respectively.}
\label{fig-adiab}
\end{figure*}

The dynamical evolution of the Saturn system under the effects of tidal forces is relevant for constraining the past evolution of the moons. In particular, the proximity of the inner evection to the planet and to the few co-orbital satellites of the system raises the question about the possible effect of this resonance on a pair of trojan moons that evolve by tidal effects.

Short and long tidal migration rates of the Saturn moons have been explored in a wide variety of scenarios by \citet{Cuk_etal_2016}. The authors concluded that either the moons are significantly younger than the planet (formed only about 100 Myr ago), or their tidal evolution must be extremely slow (implying a dissipation factor for Saturn of $Q>80\,000$). 

They also concluded that the inner evection resonance is located in a chaotic region, possibly because several harmonics of their own resonance overlap, although we showed here that this is not the case for isolated moons or trojan pairs (see Figs.~\ref{fig-Frouardzoom} and \ref{fig-maptrojans-int}). \citet{Cuk_etal_2016} proposed a mechanism to form a disk of proto-moons that could produce a favorable environment for moons to be created, which increases the possibility that some of them might harbor trojan companions. 

The dissipation factor of Saturn has traditionally been estimated to be $Q\sim 18\,000$, which gives a very low tidal migration rate. With this value, Enceladus, Tethys, Dione, and Rhea are not expected to have migrated significantly over the age of the solar system; they were probably formed at their current locations. However, based on an analysis of astrometric positions from old photographic plates and images of the Hubble Space Telescope, \citet{Lainey_etal_2012} revealed that the Saturn moons did migrate over the last century. They estimated a dissipation factor ten times smaller, $Q \sim 1\,682\pm 540$. Based on this result, \cite{Crida_Charnoz_2014} suggested that the inner moons, up to Rhea, could have been formed from the primordial spreading ring of Saturn, and in this scenario, it may be expected that many primordial trojan pairs have crossed the evection resonance.

Here, we study the effect of tidal forces in the evolution of hypothetical trojan moons by incorporating a weak-friction equilibrium tidal model into our N-body code. The tidal forces were modeled using the {closed} formulas of \citet{mignard_1979} and following the approach of \citet{Rodriguez_etal_2013} for co-orbital companions. For Saturn, we assumed a Love number $k_2 = 0.39$ and $Q = 1\,682$ \citep{Lainey_etal_2012}, and for the satellites, we assumed $k_2 = 0.05$ and $20 \leq Q \leq 200$ \citep{Cuk_etal_2016}. 

Numerical integrations using the nominal values of the system parameters are impracticable. Therefore, we scaled the total time-span and the tidal parameters down by the same factor $\alpha > 1$ to reduce the total integration time, as suggested by \citet{Zoppetti+2018}. We considered values of $\alpha$ low enough to keep the accelerated tidal evolution adiabatic with respect to the resonant and secular gravitational timescales. The results shown here correspond to $\alpha = 200$, but we performed some other numerical test using lower values to check the reliability of the results. 

We know that in the adiabatic regime, the system evolves along the family of periodic orbits, thus we ran some simulations with a single satellite. We considered a fictitious Dione-like moon that was captured in the evection resonance at $a=0.00326$~au, with $\psi$ librating around $\pm 90^\circ$ , as predicted by the analytical model. When we increased the value of $\alpha$, the libration amplitude increased and the satellite was able to cross the evection resonance without any effect on it. 

Then we ran experiments considering a pair of trojan moons with mass ratios of $m_2/m_1=2\times10^{-5}$, $m_2/m_1=1$, and $m_2/m_1=2\times10^{5}$. We recall that the first ratio  approximately corresponds to the Dione-Helene system. In all cases, the initial semi-major axis was set to $a=0.00321$~au, slightly inside of the evection resonance, and the initial eccentricity was set to the current values of Dione and Helene (0.0022). Typical results are shown in Fig.~\ref{fig-adiab}, where the evolution of $m_1$ is identify in magenta and that of $m_2$ in green. The resonant capture is observed in the bottom panels, where both evection angles $\psi_1,\psi_2$ start to librate. {For very low or very high mass ratios, the most massive component is always captured around $90^\circ$ or $270^\circ$, while the less massive component is captured around $\pm 60^\circ$ of those values. For equal masses, the components are captured around $60^\circ$ (or $140^\circ$) and $120^\circ$ (or $300^\circ$), respectively, as predicted by the analytical model. The evolution of the trojan pair inside the evection resonance implies that the eccentricities start to increase, moving along the family of periodic orbits with oscillations of higher amplitude for the less massive component. We also note that the ACR equilibrium configuration is preserved during the evolution inside the evection resonance. For low mass ratios of the trojan pair, excitation of the eccentricities contributes to destabilize the system and may lead to the loss of the less massive component, either by hitting the planet or the other component.}

{The results are similar over the whole range of values of $Q$ we tested for the satellites. On the other hand, when we considered lower values of $Q$ for Saturn ($Q<1\,000$), which implies abandoning the adiabatic regime, the evection resonance causes a small jump in the eccentricities of the trojan satellites that breaks the libration of $\Delta\varpi$, causing the system to depart from the ACR solution.} After this, the less massive satellite is captured with $\psi_i$ librating around $\pm 90^\circ$, the angle $\Delta\varpi$ starts to circulate, and $\sigma\rightarrow 180^\circ$ , leading the pair into a horseshoe regime. The system survives under this condition until a collision occurs after $\sim 10^6$~yr or sooner. 

\section{Conclusions}\label{sec.conclusions}

We have addressed the dynamics of the evection resonance in the case of co-orbital motion, and focused on trojan pairs in the ACR configuration ($\Delta\lambda=60^\circ ,\,\Delta\varpi=60^\circ$). We applied an expansion of the averaged Hamiltonian describing the motion of two bodies in trojan orbits, perturbed by a third distant body. The expansion combines the model developed by \citet{Frouard_2010} for the evection of single bodies and the model described by \citet{Robutel_2013} for the motion of trojan bodies. This model can be applied for trojan satellites in the solar system as well as for circumstellar trojan planets in binary systems. Here we consider the Saturn system. We also performed a series of numerical simulations, both to validate the predictions of the model and to better assess the actual dynamics of trojan satellites around Saturn.
Our conclusions are summarized below.
\begin{itemize}
\item Our model addressed the two types of evection resonance that may appear: the inner evection resonance, close to the planet and related to the precession of the periastron forced by the oblateness of the central body, and the outer evection resonance, far away from the planet and related to the precession of the periastron forced by the distant perturber.

\item The global structure of the phase space of the evection resonances for trojan satellites is similar to that of a single satellite, differing in that the libration centers are displaced from their standard positions ($\pm 90^\circ$ for the inner and $0^\circ, 180^\circ$ for the outer evection) by an angle that depends on the periastron difference $\varpi_2-\varpi_1$ and on the mass ratio $m_2/m_1$ of the trojan pair.

\item The distance from the planet at which the inner evection resonance occurs is independent of the mass ratio of the trojan pair. The resonance only moves toward slightly smaller distances when the obliquity of the planet is taken into account.

\item {In the Saturn system, the evolution of trojan moons into the inner evection resonance, located at $\sim 8\,R_\mathrm{S}$, is possible. Capture in this resonance can be driven by adiabatic tidal migration, exciting the eccentricities of the two trojan components. This in turn may destabilize the resonant system, leading to the loss of one of the components.}

\item The location of the outer evection resonance depends, in principle, on the mass ratio of the trojan pair, moving to longer distances as $m_2/m_1\rightarrow 1$. For the Saturn system, this would occur at distances $>0.4\,R_\mathrm{Hill}$.

\item Nevertheless, the outer evection in the Saturn system cannot exist at all for trojan moons because trojan configurations are strongly unstable at distances from Saturn longer than $\sim 0.15\,R_\mathrm{Hill}$. A pair of moons forced into a trojan configuration in the region of the outer evection always ends in a non-co-orbital configuration, with one of the moons captured in a stable orbit inside the evection resonance, while the other moon becomes unstable and is eventually lost.
\end{itemize}

In the Saturn system, the moons that are closer to the current location of the inner evection are Tethys, Dione, and Rhea. Of these, only the first two have trojan companions and are located at smaller distances from Saturn than the evection. Rhea has no trojans and is located slightly beyond the evection. If Rhea had trojan companions in the past, it may have lost them when it crossed the evection resonance. Thus, it would be interesting to study the possibility that the craters on the surface of Rhea were produced by collisions with former trojan companions that were destabilized by the evection. This study might be extended even for other regular satellites such as Titan, Hyperion, and Iapetus.

{The high eccentricity of Titan, Hyperion, and Iapetus might be related to the excitation caused by the crossing of the inner evection resonance. The eccentricity of Rhea, on the other hand, may have been excited by the evection and may have later been damped to the current value by tidal dissipation.} 

{On the other hand, the inner evection resonance moved in the past from its current location as a result of changes in the shape or obliquity of Saturn and because Saturn was closer to the Sun than it is today (e.g., \citealp{Nesvorny_Morbi_2012}). Then, Tethys and Dione might have crossed the resonance in the past and  preserved their trojan companions. This might impose constraints on the dynamical processes that led to this type of evolution.}

As for the outer evection, none of the current Saturn moons (not even the irregular moons) reach the current domain of this resonance. However, the resonance may have been relevant for the early evolution of the moons, when Saturn was closer to the Sun.

\appendix
\section{Coefficients of $\bar{\mathcal{H}}$}
We provide here the explicit expressions of the coefficient functions of the Hamiltonian Eq. (\ref{eq:hevec}) in terms of non-canonical orbital elements.
\begin{align}
A_{0i} & =-Gm_{3}m_{i}a_{i}^{2}\left(1-\nu_{i}\right)\frac{1}{4}\nonumber\\
A_{1i} & =Gm_{3}m_{i}a_{i}^{3}\frac{5}{16}\nonumber\\
C_{0} & = Gm_{3}\bar{\nu}a_{1}a_{2}\frac{1}{2}\cos2\theta_{1}\nonumber\\
C_{1} & = Gm_{3}\bar{\nu}a_{1}a_{2}\frac{1}{8}\cos4\theta_{1}\nonumber\\
C_{2} & = Gm_{3}\bar{\nu}a_{1}a_{2}\frac{9}{8}\nonumber\\
C_{3} & = Gm_{3}\bar{\nu}a_{1}a_{2}\frac{1}{8}\sin4\theta_{1}\nonumber\\
C_{4} & = Gm_{3}\bar{\nu}a_{1}a_{2}\frac{3}{16}\sin2\theta_{1}\nonumber\\
D_{i} & = -Gm_{0}m_{i}\frac{1}{2}J_{2}\frac{R_{0}^{2}}{a_{i}^{3}}\nonumber\\
\Delta_{12} & = \left(a_{1}^{2}+a_{2}^{2}-2a_{1}a_{2}\cos2\theta_{1}\right)^{1/2}\nonumber\\
B_{0} & = -\sum_{i=1}^{2}\frac{\beta_{i}\mu_{i}}{2a_{i}}\frac{2\bar{\nu}^{2}}{\mu_{1}\mu_{2}}-\frac{Gm_{1}m_{2}}{\Delta_{12}}\nonumber\\
 &  +\bar{\nu}\left(1+\frac{2\bar{\nu}}{m_{0}}\right)\sqrt{\frac{\beta_{1}\beta_{2}}{a_{1}a_{2}}}\cos2\theta_{1}\nonumber\\
B_{1} & = Gm_{1}m_{2}\left[\frac{a_{1}a_{2}(a_{1}^{2}+a_{2}^{2})\cos2\theta_{1}}{2\Delta_{12}^{5}}\right.\nonumber\\
 & \qquad\qquad\qquad-\left.\frac{a_{1}^{2}a_{2}^{2}\left(13-5\cos4\theta_{1}\right)}{8\Delta_{12}^{5}}\right]\nonumber\\
 & -\bar{\nu}\left(1+\frac{2\bar{\nu}}{m_{0}}\right)\sqrt{\frac{\beta_{1}\beta_{2}}{a_{1}a_{2}}}\frac{1}{2}\cos2\theta_{1}\nonumber\\
B_{2} & = -Gm_{1}m_{2}\left[\frac{a_{1}a_{2}(a_{1}^{2}+a_{2}^{2})\cos4\theta_{1}}{\Delta_{12}^{5}}\right.\nonumber\\
 & \qquad\qquad\qquad+\left.\frac{a_{1}^{2}a_{2}^{2}\left(\cos6\theta_{1}-17\cos2\theta_{1}\right)}{8\Delta_{12}^{5}}\right]\nonumber\\
 & +\bar{\nu}\left(1+\frac{2\bar{\nu}}{m_{0}}\right)\sqrt{\frac{\beta_{1}\beta_{2}}{a_{1}a_{2}}}\cos4\theta_{1}\nonumber\\
B_{3} & = -Gm_{1}m_{2}\left[\frac{a_{1}a_{2}(a_{1}^{2}+a_{2}^{2})\sin4\theta_{1}}{\Delta_{12}^{5}}\right.\nonumber\\
 & \qquad\qquad\qquad+\left.\frac{a_{1}^{2}a_{2}^{2}\left(\sin6\theta_{1}-35\sin2\theta_{1}\right)}{8\Delta_{12}^{5}}\right]\nonumber\\
 & +\bar{\nu}\left(1+\frac{2\bar{\nu}}{m_{0}}\right)\sqrt{\frac{\beta_{1}\beta_{2}}{a_{1}a_{2}}}\sin4\theta_{1}\nonumber
\end{align}

\begin{acknowledgements}
The authors acknowledge useful discussions with C. Beaug\'e that helped us to improve the speed of the numerical integrations with the Bulirsh-Stoer method. {We also wish to thank the anonymous referee for their comments.} This work has been supported by the National Council of Research of Argentina (CONICET) and the Rio de Janeiro State Science Foundation (FAPERJ), in the framework of a FAPERJ/CONICET bilateral project, and by the Brazilian National Council of Research (CNPq). Computations were performed at the BlaFis cluster of the University of Aveiro, Portugal, and at the SDumont cluster of the Brazilian System of HPC (SINAPAD).
\end{acknowledgements}

\bibliographystyle{aa}
\bibliography{library}

\begin{thebibliography}{33}
\expandafter\ifx\csname natexlab\endcsname\relax\def\natexlab#1{#1}\fi

\bibitem[{{Andrade-Ines} \& {Robutel}(2018)}]{Andrade-Ines_2018}
{Andrade-Ines}, E. \& {Robutel}, P. 2018, Celestial Mechanics and Dynamical
  Astronomy, 130, \#6

\bibitem[{{Beaug{\'e}} \& {Roig}(2001)}]{beauge_roig_2001}
{Beaug{\'e}}, C. \& {Roig}, F. 2001, \icarus, 153, 391

\bibitem[{{Brouwer} \& {Clemence}(1961)}]{brouwer_clemence_1961}
{Brouwer}, D. \& {Clemence}, G.~M. 1961, {Methods of celestial mechanics}
  (Academic Press, New York)

\bibitem[{Chambers(2010)}]{Chambers2010}
Chambers, J.~E. 2010, N-Body Integrators for Planets in Binary Star Systems,
  ed. N.~Haghighipour (Dordrecht: Springer Netherlands), 239--263

\bibitem[{{Cincotta} \& {Sim{\'o}}(2000)}]{Cincotta_2000}
{Cincotta}, P.~M. \& {Sim{\'o}}, C. 2000, \aaps, 147, 205

\bibitem[{{Crida} \& {Charnoz}(2014)}]{Crida_Charnoz_2014}
{Crida}, A. \& {Charnoz}, S. 2014, in IAU Symposium, Vol. 310, Complex
  Planetary Systems, Proceedings of the International Astronomical Union,
  182--189

\bibitem[{{{\'C}uk} {et~al.}(2016){{\'C}uk}, {Dones}, \&
  {Nesvorn{\'y}}}]{Cuk_etal_2016}
{{\'C}uk}, M., {Dones}, L., \& {Nesvorn{\'y}}, D. 2016, \apj, 820, 97

\bibitem[{{{\'C}uk} \& {Gladman}(2009)}]{Cuk_2009}
{{\'C}uk}, M. \& {Gladman}, B.~J. 2009, \icarus, 199, 237

\bibitem[{{Frouard} {et~al.}(2010){Frouard}, {Fouchard}, \&
  {Vienne}}]{Frouard_2010}
{Frouard}, J., {Fouchard}, M., \& {Vienne}, A. 2010, \aap, 515, A54

\bibitem[{{Giuppone} {et~al.}(2010){Giuppone}, {Beaug{\'e}}, {Michtchenko}, \&
  {Ferraz-Mello}}]{giuppone_etal_2010}
{Giuppone}, C.~A., {Beaug{\'e}}, C., {Michtchenko}, T.~A., \& {Ferraz-Mello},
  S. 2010, \mnras, 407, 390

\bibitem[{{Giuppone} \& {Leiva}(2016)}]{Giuppone_etal_2016}
{Giuppone}, C.~A. \& {Leiva}, A.~M. 2016, \mnras, 460, 966

\bibitem[{{Gott}(2005)}]{Gott_2005}
{Gott}, J.~Richard, I. 2005, Annals of the New York Academy of Sciences, 1065,
  325

\bibitem[{{Hamilton} \& {Krivov}(1997)}]{Hamilton_1997}
{Hamilton}, D.~P. \& {Krivov}, A.~V. 1997, \icarus, 128, 241

\bibitem[{{Henon}(1969)}]{Henon_1969}
{Henon}, M. 1969, \aap, 1, 223

\bibitem[{{Henon}(1970)}]{Henon_1970}
{Henon}, M. 1970, \aap, 9, 24

\bibitem[{{Innanen} {et~al.}(1997){Innanen}, {Zheng}, {Mikkola}, \&
  {Valtonen}}]{Innanen1997}
{Innanen}, K.~A., {Zheng}, J.~Q., {Mikkola}, S., \& {Valtonen}, M.~J. 1997,
  \aj, 113, 1915

\bibitem[{{Lainey} {et~al.}(2012){Lainey}, {Karatekin}, {Desmars}, {Charnoz},
  {Arlot}, {Emelyanov}, {Le Poncin-Lafitte}, {Mathis}, {Remus}, {Tobie}, \&
  {Zahn}}]{Lainey_etal_2012}
{Lainey}, V., {Karatekin}, {\"O}., {Desmars}, J., {et~al.} 2012, \apj, 752, 14

\bibitem[{{Li} \& {Christou}(2016)}]{Li_etal2016}
{Li}, D. \& {Christou}, A.~A. 2016, Celestial Mechanics and Dynamical
  Astronomy, 125, 133

\bibitem[{{Maffione} {et~al.}(2011){Maffione}, {Darriba}, {Cincotta}, \&
  {Giordano}}]{Maffione_etal_2011}
{Maffione}, N.~P., {Darriba}, L.~A., {Cincotta}, P.~M., \& {Giordano}, C.~M.
  2011, Celestial Mechanics and Dynamical Astronomy, 111, 285

\bibitem[{{Mignard}(1979)}]{mignard_1979}
{Mignard}, F. 1979, Moon and Planets, 20, 301

\bibitem[{{Murray} \& {Dermott}(1999)}]{murray_dermott_1999}
{Murray}, C.~D. \& {Dermott}, S.~F. 1999, {Solar System Dynamics} (Cambridge
  University Press)

\bibitem[{{Nesvorn{\'y}} {et~al.}(2003){Nesvorn{\'y}}, {Alvarellos}, {Dones},
  \& {Levison}}]{Nesvorny_etal_2003}
{Nesvorn{\'y}}, D., {Alvarellos}, J.~L.~A., {Dones}, L., \& {Levison}, H.~F.
  2003, \aj, 126, 398

\bibitem[{{Nesvorn{\'y}} \& {Morbidelli}(2012)}]{Nesvorny_Morbi_2012}
{Nesvorn{\'y}}, D. \& {Morbidelli}, A. 2012, \aj, 144, 117

\bibitem[{{Niederman} {et~al.}(2018){Niederman}, {Pousse}, \&
  {Robutel}}]{Niederman+2018}
{Niederman}, L., {Pousse}, A., \& {Robutel}, P. 2018, ArXiv e-prints,
  arXiv:1806.07262

\bibitem[{{Ramos} {et~al.}(2015){Ramos}, {Correa-Otto}, \&
  {Beaug{\'e}}}]{Ramos_2015}
{Ramos}, X.~S., {Correa-Otto}, J.~A., \& {Beaug{\'e}}, C. 2015, Celestial
  Mechanics and Dynamical Astronomy, 123, 453

\bibitem[{{Robutel} \& {Pousse}(2013)}]{Robutel_2013}
{Robutel}, P. \& {Pousse}, A. 2013, Celestial Mechanics and Dynamical
  Astronomy, 117, 17

\bibitem[{{Rodr{\'{\i}}guez} {et~al.}(2013){Rodr{\'{\i}}guez}, {Giuppone}, \&
  {Michtchenko}}]{Rodriguez_etal_2013}
{Rodr{\'{\i}}guez}, A., {Giuppone}, C.~A., \& {Michtchenko}, T.~A. 2013,
  Celestial Mechanics and Dynamical Astronomy, 117, 59

\bibitem[{{Roy}(1978)}]{Roy1978}
{Roy}, A.~E. 1978, {Orbital motion}

\bibitem[{{Saillenfest} {et~al.}(2016){Saillenfest}, {Fouchard}, {Tommei}, \&
  {Valsecchi}}]{Saillenfest_2016}
{Saillenfest}, M., {Fouchard}, M., {Tommei}, G., \& {Valsecchi}, G.~B. 2016,
  Celestial Mechanics and Dynamical Astronomy, 126, 369

\bibitem[{{Saillenfest} {et~al.}(2017){Saillenfest}, {Fouchard}, {Tommei}, \&
  {Valsecchi}}]{Saillenfest_2017}
{Saillenfest}, M., {Fouchard}, M., {Tommei}, G., \& {Valsecchi}, G.~B. 2017,
  Celestial Mechanics and Dynamical Astronomy, 127, 477

\bibitem[{{Spalding} {et~al.}(2016){Spalding}, {Batygin}, \&
  {Adams}}]{Spalding_2016}
{Spalding}, C., {Batygin}, K., \& {Adams}, F.~C. 2016, \apj, 817, 18

\bibitem[{{Yokoyama} {et~al.}(2008){Yokoyama}, {Vieira Neto}, {Winter},
  {Sanchez}, \& {de Oliveira Brasil}}]{Yokoyama_etal_2008}
{Yokoyama}, T., {Vieira Neto}, E., {Winter}, O.~C., {Sanchez}, D.~M., \& {de
  Oliveira Brasil}, P.~I. 2008, Mathematical Problems in Engineering, 2008, 1

\bibitem[{{Zoppetti} {et~al.}(2018){Zoppetti}, {Beaug{\'e}}, \&
  {Leiva}}]{Zoppetti+2018}
{Zoppetti}, F.~A., {Beaug{\'e}}, C., \& {Leiva}, A.~M. 2018, \mnras, 955

\end{thebibliography}
\label{lastpage}
\end{document}